\documentclass[aps,onecolumn,showpacs,superscriptaddress,groupedaddress]{revtex4}
\usepackage{txfonts}  
\usepackage{graphicx}  
\usepackage{epsfig,subfigure}
\usepackage{bm}        
\usepackage{amsfonts}
\usepackage{amssymb}

\usepackage{graphicx}
\usepackage{natbib}

\usepackage{epsfig,subfigure}
\usepackage{multirow}
\usepackage{caption}
\usepackage[dvips]{color}

\begin{document}



\title[Predictions of canonical wall bounded turbulent flows via a modified $k-\omega$ equation]
{Predictions of canonical wall bounded turbulent flows via a modified $k-\omega$ equation}

\author{Xi Chen}
\affiliation{State Key Laboratory for Turbulence and Complex Systems
and Department of Mechanics, College of
Engineering, Peking University, Beijing 100871, China}
\affiliation{Department of Mechanical Engineering, Texas Tech University, Lubbock, Texas, 79409-1021, USA}
\author{Fazle Hussain}
\affiliation{State Key Laboratory for Turbulence and Complex Systems
and Department of Mechanics, College of
Engineering, Peking University, Beijing 100871, China}
\affiliation{Department of Mechanical Engineering, Texas Tech University, Lubbock, Texas, 79409-1021, USA}
\author{Zhen-Su She}
\email{she@pku.edu.cn}
\affiliation{State Key Laboratory for Turbulence and Complex Systems
and Department of Mechanics, College of
Engineering, Peking University, Beijing 100871, China}
\date{\today}

\begin{abstract}
A major challenge in computation of engineering flows is to derive
and improve turbulence models built on turbulence physics.
Here, we present a physics-based modified $k-\omega$ equation for canonical wall bounded turbulent flows (boundary layer, channel and pipe), predicting both mean velocity profile (MVP)
and streamwise mean kinetic energy profile (SMKP) with high accuracy over a wide range of Reynolds number ($Re$).
The result builds on a multi-layer quantification of wall flows, which allows a significant modification of the $k-\omega$ equation \cite{wilcox}. Three innovations are introduced:
First, an adjustment of the Karman constant to 0.45 is set for the overlap region with a logarithmic MVP.
Second, a wake parameter models the turbulent transport near the centerline.
Third, an anomalous dissipation factor represents the effect of a meso layer in the overlap region.
Then, a highly accurate (above 99\%) prediction of MVPs is obtained in Princeton pipes \citep{zagarola1998,Mckeon2004},
improving the original model prediction by up to 10\%. Moreover, the entire SMKP, including the newly observed outer
peak \citep{Hultmark2012}, is predicted. With a slight change of the wake parameter, the model also yields accurate predictions for channels and boundary layers.
\end{abstract}



\maketitle

\section{Introduction}
Prediction of canonical wall-bounded turbulent flows are of great importance for both theoretical
study and engineering applications \citep{smits2013}. They are the fundamental problems of turbulence often benchmarked to validate computational fluid dynamics (CFD) models. One example is the celebrated log-law, an essential part of the Reynolds averaged Navier-Stokes (RANS)
simulations of engineering flows \citep{pope2000turbulent,Davidson2004}. Despite extensive studies over several decades, simultaneously accurate predictions of
both the mean velocity and fluctuation intensities still remain as an outstanding problem \citep{marusic2010wall}. One challenge concerns the universality (or not) of the Karman constant $\kappa$ \citep{Smits}.
Spalart \citep{Spalart} asserts that a difference of 6\% in $\kappa$ can change the predicted skin
friction coefficient by up to 2\% at a length Reynolds number of
$10^8$ of a typical Boeing 747. On the other hand, more than 10\% variation of
$\kappa$ is reported in the literature \citep{marusic2010wall}. The `Karman coefficient' instead of Karman constant is even suggested \cite{monkewitz2007, monkewitz2008}; yet no sign of convergence is apparent for this vivid debate \citep{Alfredsson2013,segalini2013}. In this work, we carry out a careful comparison
to illustrate what optimal $\kappa$ should be used in the $k-\omega$ model. It turns out that
$\kappa=0.45$ \citep{shenjp,wuyoups} would result in a 99\% accurate
description of more than 30 MVPs measured in channel, pipe and TBL flows over a wide range of $Re$'s.

Another challenge is the outer peak of the SMKP (or streamwise fluctuation intensity), not predicted by existing turbulence models. Some prior models (e.g. the $k-\omega$ model) assume a {Bradshaw-like} constant which yields an equilibrium state
where turbulent fluctuations reach a constant plateau in the overlap region \citep{wilcox}. However,
this is against recent accurate measurements \citep{Morrison2004,Hultmark2012,Vallikivi2015jfm}.
Unlike studies on the
MVP \citep{nickels2004, javier2006, Panton2007, lvovprl}, theoretical works on the SMKP are much fewer. The logarithmic scaling
by Townsend \cite{Townsend1976} and Perry \emph{et al} \cite{Perry1986} are only observed recently, and the composite description of SMKP through piece-wise
functions \citep{marusic2003pof, Smits2010, Alfredsson2011,Alfredsson2012,Vassilicos2015} are difficult
to incorporate in the RANS models. Hence, improvement of the SMKP prediction is more subtle.

In this paper, we present a modified $k-\omega$ model which shows the improvement of predictions of both MVP and SMKP based on the understanding of multi-layer physics of wall turbulence. The $k-\omega$ model is widely used
in engineering applications, enabling predictions of both the mean and
fluctuation velocities. Here, we introduce three modifications to the latest version of the model by Wilcox (2006) \citep{wilcox}:
1) an adjustment of the Karman constant, 2) a wake parameter to take into account the enhanced turbulent transport effect, and 3) an anomalous dissipation factor to represent a meso layer in the overlap
region. These modifications yield a more accurate (above 99\%) description of MVPs in Princeton pipe data
for a wide range of $Re$'s, improving the Wilcox $k-\omega$ model prediction by up to 10\%.
Moreover, they yield a good prediction of the entire SMKP, where the newly observed outer peak
\citep{Hultmark2012} is also captured. With a slight change of the wake parameter,
the model yields also very good predictions for turbulent channels and turbulent boundary layers.

The paper is organized as follows. Section II introduces the predictions by original $k-\omega$ equation as well as its
approximate solutions in the overlap region. In Section III, we present two of the three modifications mentioned above
and compare with empirical MVP data.
Section IV introduces the third modification for prediction of SMKP. Conclusion
and discussion are presented in Section V.

\section{The original $k-\omega$ equation's predictions}

In 1942, Kolmogorov \cite{kolmogorov1942} introduced a closure for turbulent kinetic energy $k$ through the dissipation rate $\epsilon$, via $\omega=\epsilon/k$. As the dimension of $\epsilon$ is $[\ell]^2[t]^{-3}$, the dimension for $\omega$ is $[t]^{-1}$,
which is interpreted as the energy cascade rate. Kolmogorov suggested a transport equation
for $\omega$, parallel to that for $k$. Independent of Kolmogorov, Saffman \cite{Saffman1970} also formulated a $k-\omega$ model, and since then, a further development and application of $k-\omega$ model has been pursued \citep{Saffman1974,wilcox1988}.
The latest version established by Wilcox \citep{wilcox} reads
\begin{eqnarray}\label{eq:ko1a}
\frac{{\partial k}}{{\partial t}} + {U_j}\frac{{\partial k}}{{\partial {x_j}}} &=& {\tau _{ij}}\frac{{\partial {U_i}}}{{\partial {x_j}}}
- {\beta ^*}k\omega  + \frac{\partial }{{\partial {x_j}}}\left[ {(\nu  + {\sigma ^*}\nu_T){\frac{\partial k}{\partial {x_j}}}} \right]\\
\frac{{\partial \omega }}{{\partial t}} + {U_j}\frac{{\partial \omega }}{{\partial {x_j}}} &=& \alpha \frac{\omega }{k}
{\tau _{ij}}\frac{{\partial {U_i}}}{{\partial {x_j}}} - \beta {\omega ^2} + \frac{\partial }{{\partial {x_j}}}
\left[ {(\nu  + \sigma \nu_T)\frac{{\partial \omega }}{{\partial {x_j}}}} \right] \nonumber\\&+& \frac{{{\sigma _d}}}
{\omega }\frac{{\partial k}}{{\partial {x_j}}}\frac{{\partial \omega }}{{\partial {x_j}}}.\label{eq:ko1b}
\end{eqnarray}
In (\ref{eq:ko1a}) and (\ref{eq:ko1b}), the right hand side (r.h.s.) are production, dissipation,
viscous diffusion and turbulent transport in order, and there is an additional cross-diffusion term
in (\ref{eq:ko1b}). In its original version, $k$ denotes the total kinetic energy - sum of streamwise, wall normal and spanwise
components. As noted by Wilcox \cite{wilcox1988}, it is not critically important
whether $k$ is taken to be the full kinetic energy of {three} components or, alternatively,
the streamwise component only. The rationale is that in both cases, the quantity $k$ represents
a measure of fluctuation intensity. In what follows, we assume that $k=\langle u'u'\rangle/2$, because in parallel flows such as
in pipe or in channel, (\ref{eq:ko1a}) is more similar to the budget equation of $\langle u'u'\rangle/2$, {in which the production} is by mean shear, in contrast to
the spanwise and normal fluctuation intensities by pressure transports.
As we show below, our definition of $k$ with appropriate modifications (for turbulent transports) yields good predictions
to recent experiments of pipe and channel.

Consider the fully developed channel flow, for example.
The left hand side (l.h.s.) of (\ref{eq:ko1a}) and (\ref{eq:ko1b}) are all zero. The two equations are
coupled with the streamwise mean momentum equation (mean velocities in the wall normal $y$
and spanwise $z$ directions are zero), which, after integration once in $y$, reads:
\begin{equation}
\nu \partial U/\partial y - \langle u'v'\rangle=u_\tau^2 (1-y/\delta),
\end{equation}
where $u_\tau=\sqrt{\nu \partial U/\partial y|_{y=0}}$ is the friction velocity, $\delta$ is the channel half
width. For convenience, $y$ will designate $y/\delta$ in the remainder of this paper unless otherwise specified.
Introducing the viscous normalization (using velocity scale $u_\tau$ and length scale $\nu /u_\tau$) indicated by
super script $+$, all above equations are nondimensionalized as:
\begin{eqnarray}
\quad {S^ + }{W^ + } - {\beta ^ * }{k^ + }{\omega ^ + } + \frac{d}{{d{y^ + }}} \left[ (1 + {\sigma ^*}\nu_T^ +)
\frac{{d{k^ + }}}{{d{y^ + }}} \right]   &=& 0\label{eq:Kplus}\\
\alpha  \frac{{{\omega^ + }}}{{{ k^ + }}} {S^ + }{W^ + } - \beta {\omega ^ {+2} } + \frac{d}{{d{y^ + }}}
\left[ (1 + {\sigma }\nu_T^ +) \frac{{d{\omega^ + }}}{{d{y^ + }}} \right] \nonumber\\+
\frac{{{\sigma _d}}}{\omega^+ }\frac{{d k^+}}{{d {y^+}}}\frac{{d \omega^+ }}{{d {y^+}}} &=& 0\label{eq:omegaplus}\\
{S^ + } + {W^ + } - 1 + {{{y^ + }} \mathord{\left/
 {\vphantom {{{y^ + }} {{\rm{R}}{{\rm{e}}_\tau }}}} \right.
 \kern-\nulldelimiterspace} {{\rm{R}}{{\rm{e}}_\tau }}}&=&0,\label{eq:MME}
\end{eqnarray}
where $S^+=d U^+/d y^+$ (we change partial derivative to ordinary derivative since only $y$ dependence is considered here);
$W^+=-\langle u'v'\rangle^+$; and $1-y^+/Re_\tau=r$ is the normalized distance from the centerline.

Note that for pipes, cylindrical coordinates are more convenient.
In this case, the mean momentum equation is the same as (\ref{eq:MME}) (see appendix A), but the $k$ and $\omega$ equations are:
\begin{eqnarray}
\quad {S^ + }{W^ + } - {\beta ^ * }{k^ + }{\omega ^ + } +\frac{1}{r}\frac{d}{{d{y^ + }}} \left[ r(1 + {\sigma^* }\nu_T^ +)
\frac{{d{k^ + }}}{{d{y^ + }}} \right] &=& 0\label{eq:Kpluspipe}\\
\alpha  \frac{{{\omega^ + }}}{{{ k^ + }}} {S^ + }{W^ + }- \beta {\omega ^ {+2} } +\frac{1}{r} \frac{d}{{d{y^ + }}}
\left[ r(1 + {\sigma }\nu_T^ +) \frac{{d{\omega^ + }}}{{d{y^ + }}} \right]\nonumber\\+ \frac{{{\sigma _d}}}{\omega^+ }
\frac{{d k^+}}{{d {y^+}}}\frac{{d \omega^+ }}{{d {y^+}}} &=& 0.\label{eq:omegapluspipe}
\end{eqnarray}
Only the diffusion and transport terms are modified in pipes compared to channels,
introducing a slightly higher mean velocity in the bulk of pipes at the same $Re_\tau$.

The mean quantities of a TBL follow a two dimensional equation with a streamwise development. However, in this paper,
we only consider the description of
the streamwise MVP and SMKP for a given $Re_\tau$. For the $k-\omega$ model of TBL, the Cartesian coordinates are
used as in channels.

Note that the $k-\omega$ equation defines the eddy viscosity as
\begin{equation}\label{eq:eddyv}
\nu _T^ +\equiv {{{W^ + }} \mathord{\left/
 {\vphantom {{{W^ + }} {{S^ + }}}} \right.
 \kern-\nulldelimiterspace} {{S^ + }}}=  {{{\alpha ^ * }{k^ + }} \mathord{\left/
 {\vphantom {{{\alpha ^ * }{k^ + }} {{\omega ^ + }}}} \right.
 \kern-\nulldelimiterspace} {{\omega ^ + }}}.
\end{equation}
Thus, (\ref{eq:Kplus})-(\ref{eq:eddyv}) are a set of closed equations to predict both MVP and SMKP.
The (complicated) parameter setting in the original $k-\omega$ equation \citep{wilcox} is:
${\alpha ^ * } = \frac{{\alpha _0^ *  + {{{k^ + }}
\mathord{\left/
 {\vphantom {{{k^ + }} {\left( {{R_k}{\omega ^ + }} \right)}}} \right.
 \kern-\nulldelimiterspace} {\left( {{R_k}{\omega ^ + }} \right)}}}}{{1 + {{{k^ + }} \mathord{\left/
 {\vphantom {{{k^ + }} {\left( {{R_k}{\omega ^ + }} \right)}}} \right.
 \kern-\nulldelimiterspace} {\left( {{R_k}{\omega ^ + }} \right)}}}}$, $\alpha  = \frac{{{\alpha _\infty }}}{{{\alpha ^ * }}}\frac{{{\alpha _0} + {{{k^ + }} \mathord{\left/
 {\vphantom {{{k^ + }} {\left( {{R_\omega }{\omega ^ + }} \right)}}} \right.
 \kern-\nulldelimiterspace} {\left( {{R_\omega }{\omega ^ + }} \right)}}}}{{1 + {{{k^ + }} \mathord{\left/
 {\vphantom {{{k^ + }} {\left( {{R_\omega }{\omega ^ + }} \right)}}} \right.
 \kern-\nulldelimiterspace} {\left( {{R_\omega }{\omega ^ + }} \right)}}}}$ and ${\beta ^ * } = \beta _0^ * \frac{{{{100{\beta _0}} \mathord{\left/
 {\vphantom {{100{\beta _0}} {27}}} \right.
 \kern-\nulldelimiterspace} {27}} + {{\left[ {{{{k^ + }} \mathord{\left/
 {\vphantom {{{k^ + }} {\left( {{R_\beta }{\omega ^ + }} \right)}}} \right.
 \kern-\nulldelimiterspace} {\left( {{R_\beta }{\omega ^ + }} \right)}}} \right]}^4}}}{{1 + {{\left[ {{{{k^ + }} \mathord{\left/
 {\vphantom {{{k^ + }} {\left( {{R_\beta }{\omega ^ + }} \right)}}} \right.
 \kern-\nulldelimiterspace} {\left( {{R_\beta }{\omega ^ + }} \right)}}} \right]}^4}}}$ are functions of $y^+$, with ${R_k} = 6$, ${R_\omega } = 2.61$ and ${R_\beta } = 8$ indicating the specific transition locations of $\alpha^\ast$, $\alpha$ and $\beta^\ast$, respectively. Other coefficients are constants, i.e. ${\alpha _\infty } = 0.52$; $\alpha _0^ *  = {{{\beta _0}} \mathord{\left/
 {\vphantom {{{\beta _0}} 3}} \right.
 \kern-\nulldelimiterspace} 3}$; ${\alpha _0} = {1 \mathord{\left/
 {\vphantom {1 9}} \right.
 \kern-\nulldelimiterspace} 9}$; $\beta _0^ *  = 0.09$; $\beta  = {\beta _0} = 0.0708$; $\sigma  = 0.5$; ${\sigma ^ * } = 0.6$; $\sigma_d=0.125$.

\subsection{Approximate solutions in the overlap region}

Before we display the numerical result of the $k-\omega$ equation, it is important to discuss the approximate solutions in
the overlap region, in order to reveal the role of $\kappa$. These local solutions, also obtained in \cite{wilcox},
help us to understand the difference between
numerical result and empirical data, and to suggest plausible improvement, as shown below.

At first, let us present a general expression for mean shear and kinetic energy in the outer flow (including the overlap and
wake or central core regions). In this region, the mean shear $S^+$ is much smaller than the Reynolds shear stress $W^+$, and we have
$W^+\approx r$ in (\ref{eq:MME}). Therefore, $S^+$ is given as
\begin{equation}S^+=W^+/\nu_T^+\approx r/\nu_T^+\label{eq:nut}\end{equation}
In the $k$-equation (\ref{eq:ko1a}), we further introduce a ratio function between dissipation and production,
i.e. $\Theta_\nu=\varepsilon_k^+/(S^+W^+)=\beta^\ast k^+\omega^+/(S^+W^+)$, which gives the leading balance
transition. It follows that $\Theta_\nu\approx \beta^\ast k^+\omega^+ \nu_T^+/r^2$ by substituting (\ref{eq:nut}) in, which,
together with $\omega^+ \nu_T^+=\alpha^\ast k^+$, leads to the solution
\begin{equation}k^+\approx r \sqrt{\Theta_\nu/(\alpha^\ast\beta^\ast)}.\label{eq:ke}\end{equation}

Furthermore, (approximate) analytic solutions can be derived specifically in the overlap layer, i.e. for $y^+$
varying from 40 to $0.1Re_\tau$ (exact values of the bounds do not affect the analysis below). In this layer,
$W^+\approx r\approx 1$ (for large $Re_\tau$), $\alpha ^\ast \approx 1$, $\alpha\approx \alpha_\infty$,
$\beta\approx\beta_0$, $\beta^\ast\approx \beta^\ast _0$, $\Theta_\nu\approx 1$ (the quasi-balance between
production and dissipation in {(\ref{eq:Kpluspipe})}).
Therefore, (\ref{eq:ke}) tells us
\begin{equation}k^+\approx 1/\sqrt{\beta^\ast_0}\approx 3.3\label{eq:k33}\end{equation}
and hence $\langle u'u'\rangle^+= 2k^+\approx 6.6$. {Note that $k^+/W^+\approx 3.3$ is a Bradshaw-like constant.}

Moreover, (\ref{eq:nut}) yields $\nu^+_T\approx 1/S^+$; from (\ref{eq:eddyv}) and (\ref{eq:k33}),
$\omega^+\approx S^+/\sqrt{\beta_0^\ast}$. Substituting these two results into
{(\ref{eq:omegapluspipe})} yields:
\begin{equation}\label{eq:wilc}
\alpha_\infty {S^{ + 2}} - \beta_0 {S^{ + 2}}/\beta _0^* + (\sigma/\sqrt {\beta _0^*}) \frac{d}{{d{y^ + }}}(\frac{1}{{ {S^ + }}}\frac{{d{S^ + }}}{{d{y^ + }}}) \approx 0
\end{equation}
(note that the diffusion and stress limited terms are much smaller and hence neglected). Above equation owns an analytical solution
\begin{equation}\label{eq:log}
S^+\approx 1/(\kappa y^+)   \quad\quad \Longrightarrow  \quad\quad U^+\approx \kappa^{-1} \ln y^+ +B
\end{equation}
with the Karman constant determined by four model coefficients:
\begin{equation}\label{eq:kappa}
  \kappa  = \sqrt {\frac{{\beta_0  - \alpha_\infty {\beta_0 ^ * }}}{{\sigma \sqrt {{\beta_0 ^ * }} }}}
\end{equation}
With ${\alpha _\infty } = 0.52$, ${\beta _0} =
0.0708$, $\beta _0^ *  = 0.09$, $\sigma  = 0.5$ as
Wilcox originally suggested, the resulting $\kappa$ in (\ref{eq:kappa})
is 0.40. Also note that the log-law additive
constant $B$ is about 5.6 if $\kappa = 0.40$.

Therefore, $k-\omega$ ensures a log-profile for the mean velocity and a constant kinetic energy in the overlap region. However,
these two solutions depart significantly from data (shown below), and their modifications are the main results of this paper.

\subsection{Numerical results of the original $k-\omega$ equation}

We use the the numerical code presented in \cite{wilcox} to calculate the above $k-\omega$ equation. Note that a central difference scheme is used, and the grid is set to guarantee that there are at least
ten points below $y^+=1$. A logarithmic mean velocity with a constant kinetic energy is used as the initial solution for
iteration, and the convergence condition {following that used in \cite{wilcox}} is set as $|U^{n+1}-U^{n}|\leq 1\times10^{-5}$ where $n$ denotes the
number of iterations. Below, the $k-\omega$ equation (model) indicates the Wilcox (2006) version,
while $k-\omega$ SED \citep{she2009} indicates the modified one unless otherwise specified.

\begin{figure}
\subfigure []{\includegraphics[width = 10 cm]{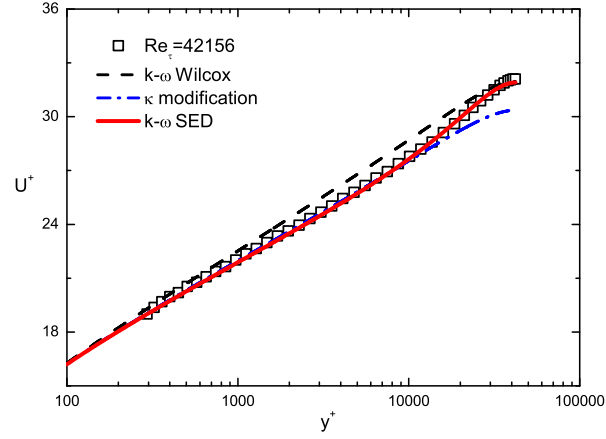}}
\subfigure []{\includegraphics[width = 10 cm]{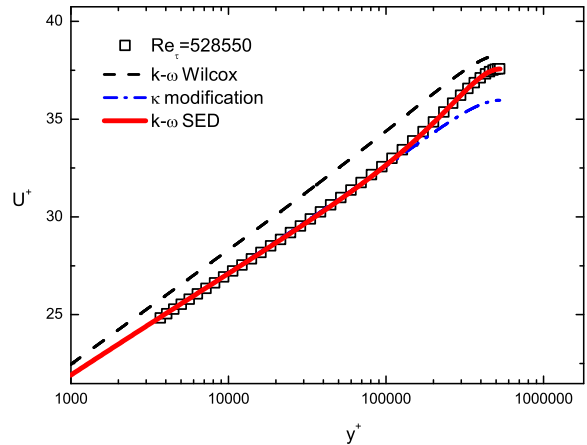}}
\caption{Comparison of the MVP prediction by the Wilcox $k-\omega $ model (dashed line) with Princeton pipe measurements \citep{zagarola1998} (symbols) at $Re_\tau=42156$ (a) and $528550$ (b). The dash dotted line indicates our prediction by changing Karman constant 0.40 into 0.45, which show clear improvement in the overlap region. The solid line is the prediction of the modified $k-\omega $ model discussed below (see Section III). } \label{fig:MVP1}
\end{figure}

The numerical solution of MVPs by the $k-\omega $ model, i.e. (\ref{eq:MME})-(\ref{eq:eddyv}),
is shown in
figure \ref{fig:MVP1}, compared with Princeton pipe data \citep{zagarola1998} at $Re_\tau=42156$ and $528550$.
Note that the deviation in the
outer region is significant, especially for $Re_\tau=528550$, with over-prediction of the mean
velocity everywhere. Note also an unphysical feature of this model:
the predicted MVP near the centerline is lower than the initial log-profile, whereas the empirical data
always show a wake structure higher than the log-profile. We thus propose two modifications
of the $k - \omega $ model: reset Karman constant to
0.45, and add a wake correction enhancing the turbulent transport, where a much better prediction is achieved (figure \ref{fig:MVP1}).

\begin{figure}
\subfigure []{\includegraphics[width = 10 cm]{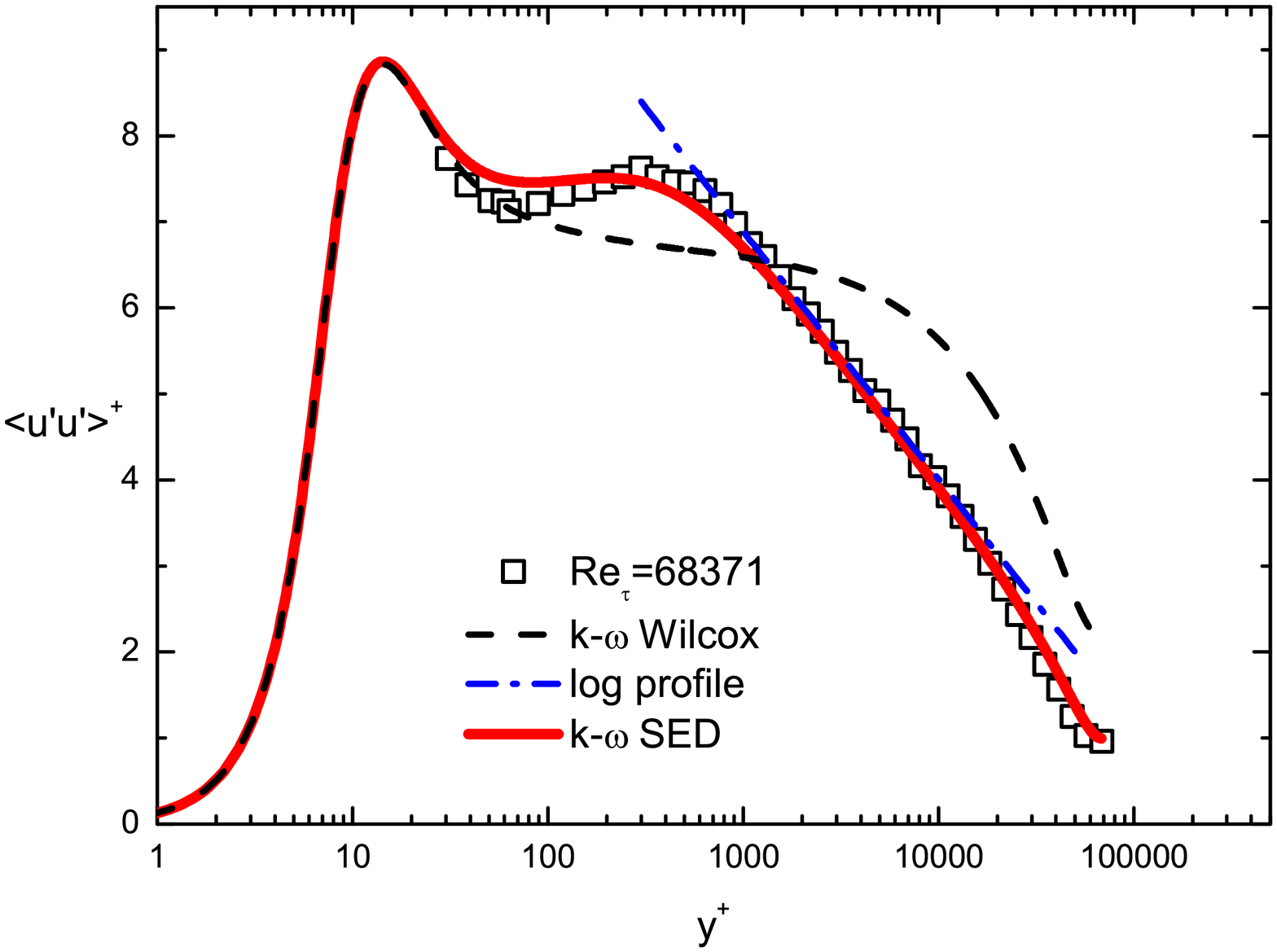}}
\subfigure []{\includegraphics[width = 10 cm]{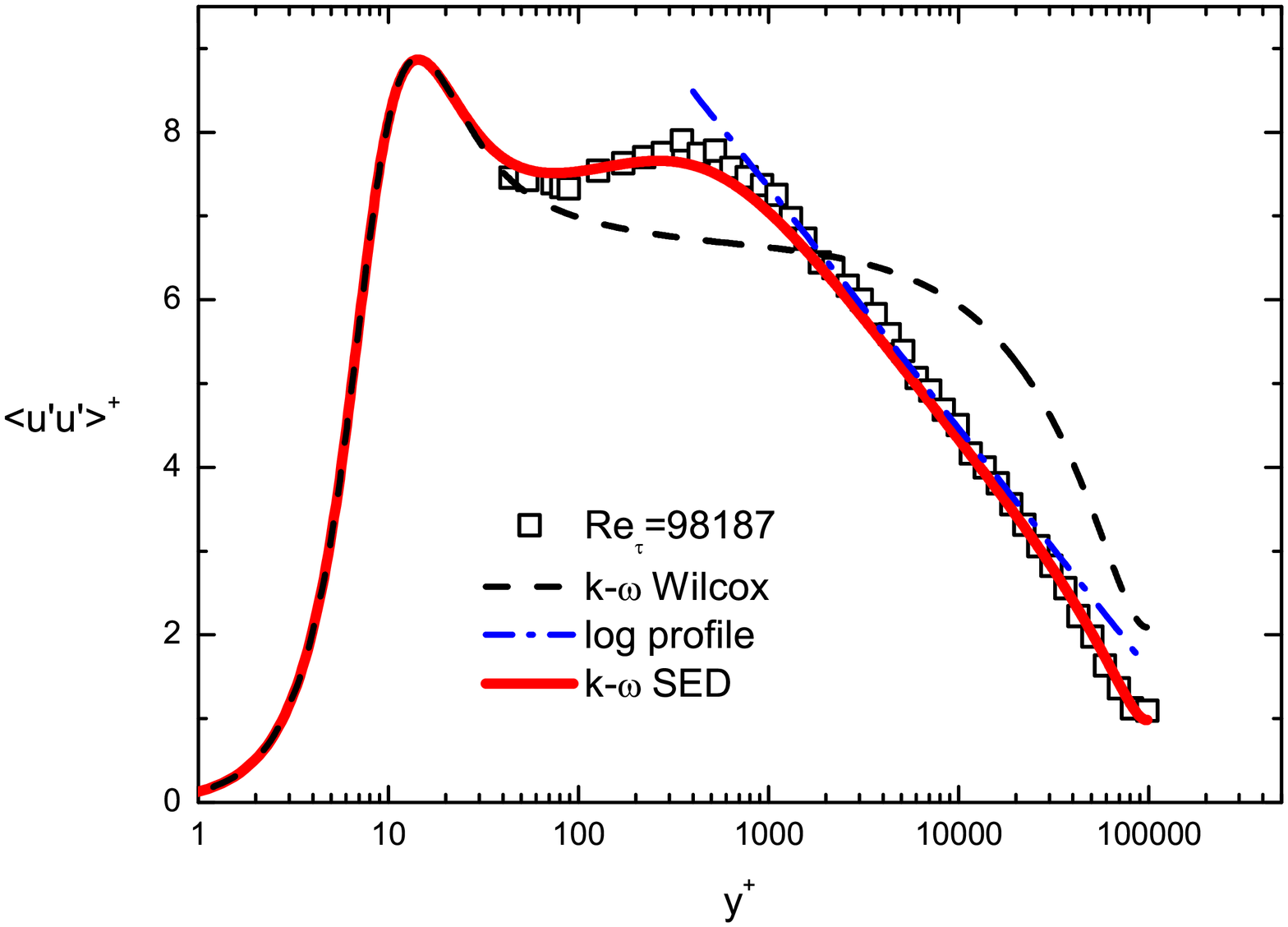}}
  \caption{SMKP prediction by the Wilcox $k-\omega $ model (dashed line) compared with Princeton pipe measurements \citep{Hultmark2012} (symbols)
  at {$Re_\tau=68371$ (a) and $98187$ (b)}.
  The dash dotted line indicates the logarithmic profile in (\ref{eq:Klog}), and the solid line is the prediction of our
  modified $k-\omega $ model discussed below (see Section IV). } \label{fig:MKP1}
\end{figure}

For SMKP, the result is shown in figure \ref{fig:MKP1}. Note that we choose two recent
profiles from Princeton pipe \citep{Hultmark2012} at $Re_\tau=68371$ and $98187$, where the outer peak and the approximate
logarithmic profile in the bulk flow region are observed, i.e.
\begin{equation}\label{eq:Klog}
\langle u'u'\rangle^+ = 2{k^ + } \approx  - 1.25\ln (y^+/Re_\tau) + 1.61.
\end{equation}
This comparison shows a severe deviation in the
whole outer region (for $y^+$ over about 40), where predicted constant profile for $\langle u'u'\rangle^+\approx 6.6$
from $y^+\approx 10^2$ to $10^4$ by (\ref{eq:k33}) is
not observed while data show a notable outer peak at $y^+\le500$ in the overlap region.
In addition, the departure of the $k-\omega$ prediction from the log profile of (\ref{eq:Klog}) is also
quite obvious, with more rapid decrease from the plateau towards the centerline. Moreover, the centerline kinetic energy is
over-predicted. Thus, a significant modification is needed to improve the prediction of SMKP.

\section{Modification of the $k - \omega $ equation for MVP}

First, we observe that the $k-\omega $ model prediction of MVP has a
higher log-linear slope ($1/\kappa$) than the
empirical data, suggesting that theoretical setting of $\kappa$ (and also $B$) is too small. Thus, our first
attempt is to increase $\kappa$ from $0.4$ to $0.45$ (according to \citep{shenjp,wuyoups});
and a higher $B$. A specific choice is to increase $\alpha_\infty$ from 0.52 to 0.57 for increasing $B$;
and to set $\sigma  = {{\left( {\beta_0  - \alpha_\infty {\beta_0 ^ * }} \right)}
\mathord{\left/
 {\vphantom {{\left( {\beta_0  - \alpha_\infty {\beta_0 ^ * }} \right)} {\left( {{\kappa ^2}\sqrt {{\beta_0 ^ * }} } \right)}}} \right.
 \kern-\nulldelimiterspace} {\left( {{\kappa ^2}\sqrt {{\beta_0 ^ * }} } \right)}} \approx 0.488$ by
following (\ref{eq:kappa}) to guarantee $\kappa  = 0.45$. This amounts to adjusting only two parameters
($\alpha _\infty$ and $\sigma$) to obtain desired $\kappa$ and $B$.
The reason we choose to adjust $\alpha _\infty$ and $\sigma $, instead of $\beta$
in (\ref{eq:kappa}), is that the latter affects both MVP and SMKP.
Indeed, this simple modification yields a very promising
result, as shown in figure \ref{fig:MVP1} (dash dotted blue lines), which significantly
improves the description of the overlap region.

Inspecting the outcome in figure \ref{fig:MVP1}, the second deficiency of the
original $k-\omega$ model
becomes more pronounced: approaching the centerline, the MVP prediction of the $k-\omega $
equation increases too slowly, even below the log-profile, contrary to the trend shown in empirical data.
To rectify this deficiency, an approximate analytic solution of the $k- \omega$
equation for the entire outer flow becomes helpful (see below).

\subsection{Outer similarity}

To rectify the deficient wake structure of the $k-\omega$ equation, the analysis of the overlap region
is too restricted, and we must derive a new set of approximate balance equations valid for the entire outer flow - including
the effects of turbulent transport which plays the dominant role in the central core region.
In fact, this is possible since we have the following analytic expression for the entire outer flow region,
$k^+\approx \nu _T^{+}\omega^+$,
${S^ + } \approx {r \mathord{\left/
 {\vphantom {r {\nu _T^ + }}} \right.
 \kern-\nulldelimiterspace} {\nu _T^ + }}$. Substituting them back into {(\ref{eq:Kpluspipe}) and (\ref{eq:omegapluspipe})},
 only neglecting the diffusion terms, we obtain:
\begin{eqnarray}\label{eq:Kout}
  {{{r^2}} \mathord{\left/
 {\vphantom {{{r^2}} {\nu _{_T}^{}}}} \right.
 \kern-\nulldelimiterspace} {\nu _{_T}^{}}} - {\beta ^*}\nu _{_T}^{}{\omega ^2} + \frac{1}{r}\frac{d}{{dr}}\left( r{\sigma ^*}{\nu _{_T}^{}\frac{{d\left( {\nu _{_T}^{}\omega } \right)}}{{dr}}} \right) = 0\\ \label{eq:Omegaout}
  \alpha {{{r^2}} \mathord{\left/
 {\vphantom {{{r^2}} {\nu _{_T}^{2}}}} \right.
 \kern-\nulldelimiterspace} {\nu _{_T}^{2}}} - \beta {\omega ^2} +  \frac{1}{r}\frac{d}{{dr}}\left( r\sigma {\nu _{_T}^{}\frac{{d\omega }}{{dr}}} \right)+\frac{{{\sigma _d}}}{\omega }\frac{{d (\nu_T \omega)}}{{d {r}}}\frac{{d \omega }}{{d {r}}} = 0
\end{eqnarray}
where $\nu _T^{} = {{\nu _T^ + } \mathord{\left/
 {\vphantom {{\nu _T^ + } {{{{\mathop{\rm Re}\nolimits} }_\tau }}}} \right.
 \kern-\nulldelimiterspace} {{{{\mathop{\rm Re}\nolimits} }_\tau }}}$ and
$\omega  = {\omega ^ + }{{\mathop{\rm Re}\nolimits} _\tau }$ are normalized to eliminate the $Re$ effect in the outer flow.
In other words, by changing the primary variables ($k^+$ and $S^+$) into $\nu_T$ and $\omega$,
we obtain a (closed) set of invariant equations for the outer flow.

\begin{figure}
\subfigure[]{\includegraphics[width = 10 cm]{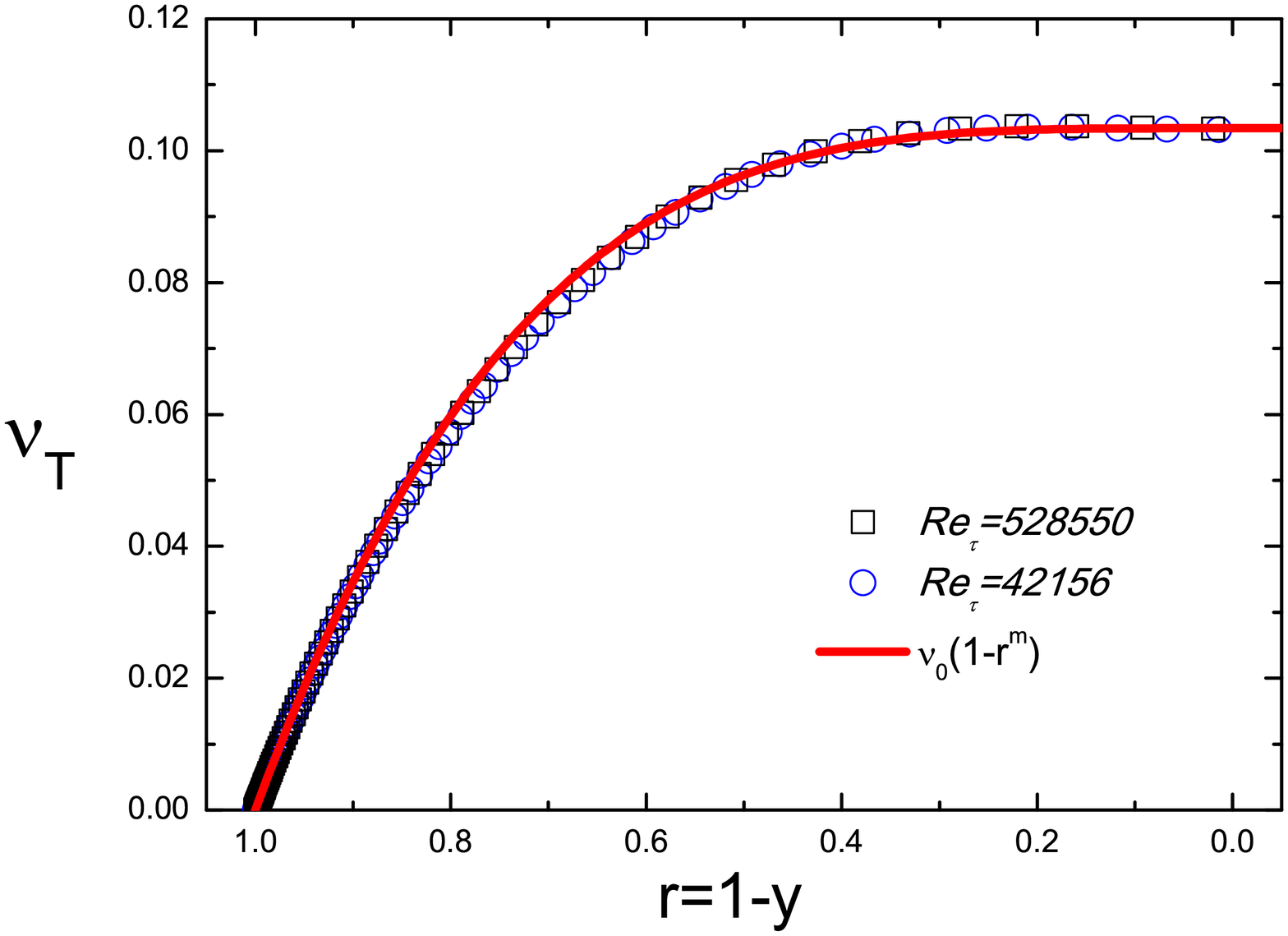}}
\subfigure[]{\includegraphics[width = 10 cm]{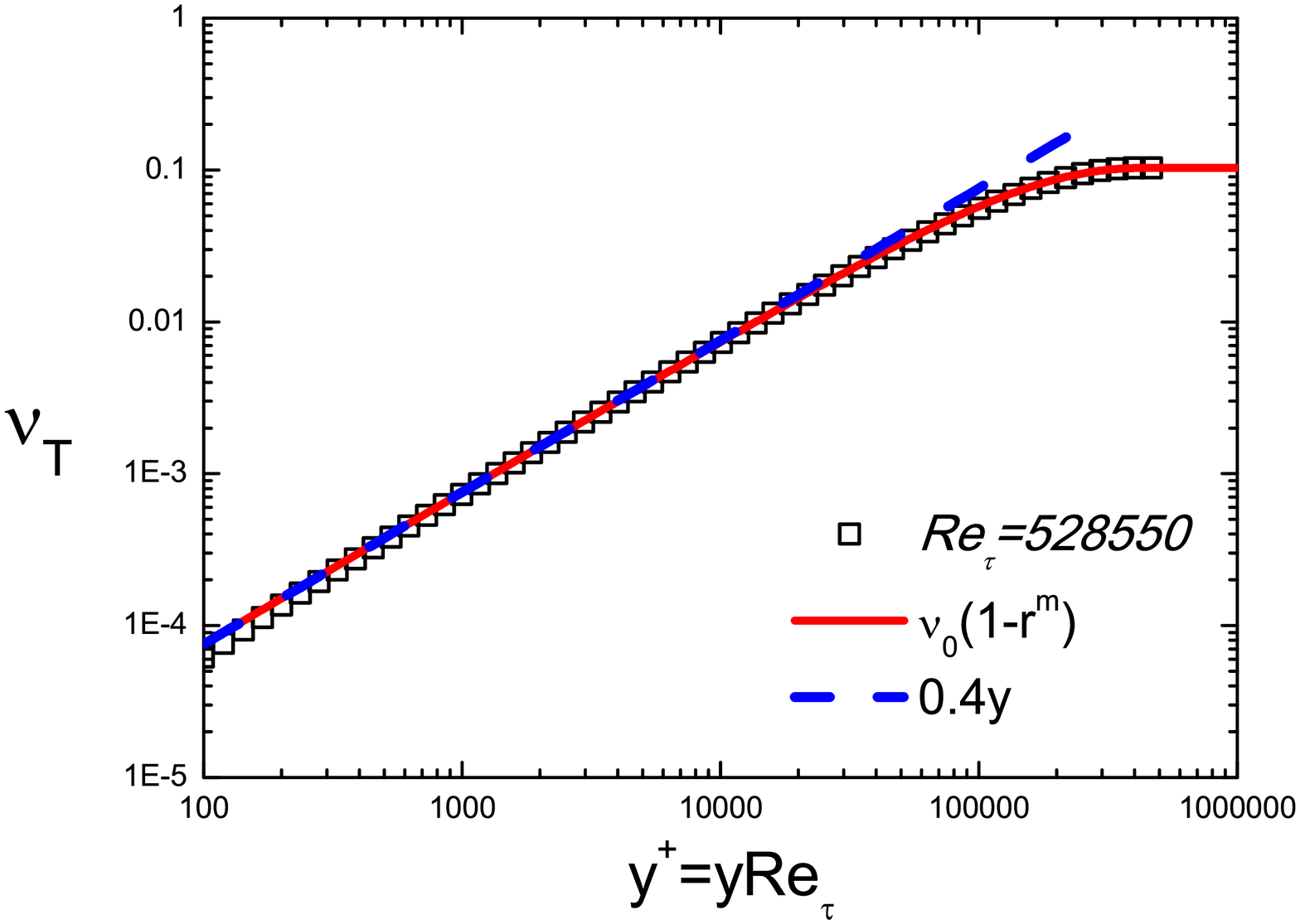}}
\caption{Validation of the bulk solution ansatz (\ref{eq:nutKO}) (solid line) for the eddy viscosity in the Wilcox $k - \omega $ model by its full numerical solution (symbols). (a) In center coordinate $r = 1 - y$. (b) In wall coordinate, ${y^ + } = y{{\mathop{\rm Re}\nolimits} _\tau }$. Note that the numerical profiles collapse for two very different \emph{Re}'s, indicating self-similarity for the eddy viscosity in the outer flow. Dashed line indicates the near wall linear scaling ${\nu _T} = \kappa y$.}
\label{fig:nut}
\end{figure}

Indeed, this self-similarity is validated as shown in figure \ref{fig:nut} which displays a full collapse of $\nu_T$ profiles
for $Re_\tau=42158$ and $Re_\tau=528550$. Note that as $\nu_T$ determines the mean shear and hence MVP, it is important to
pursue an approximate analytic description of $\nu_T$ for the entire outer flow.
It turns out that $\nu_T$ satisfies a specific ansatz
\begin{equation}\label{eq:nutKO}
  \nu _T^{k - \omega } \approx {\nu _0}\left( {1 - {r^m}}\right),
\end{equation}
explained as below. Note that (\ref{eq:nutKO}) displays two local states, i.e. $\nu_T\rightarrow \nu_0$ as $r\rightarrow 0$ and
$\nu_T\rightarrow \nu_0m(1-r)=\nu_0 m y\rightarrow 0$ as $r\rightarrow 1$ ($y=1-r\rightarrow 0$), the latter is the approximate solution in the
overlap region. Here $\nu_0$ is the centerline eddy viscosity and $m$ is the defect power law scaling exponent.
According to data in figure \ref{fig:nut}, $\nu _0 \approx 0.103$ (independent of $Re$), which determines
the scaling exponent: $m = {\kappa  \mathord{\left/
 {\vphantom {\kappa  {\nu _0^{}}}} \right.
 \kern-\nulldelimiterspace} {\nu _0^{}}} \approx {{0.40} \mathord{\left/
 {\vphantom {{0.40} {0.103}}} \right.
 \kern-\nulldelimiterspace} {0.103}} \approx 3.88$.
This amounts to a matching with the logarithmic solution ${\nu _T} = \kappa y$.
With only one empirical coefficient (i.e. $\nu_0$), (\ref{eq:nutKO}) is shown to be in
excellent agreement with the full numerical solution of the original $k-\omega$ equation, as illustrated in figure \ref{fig:nut}.
Incidentally, by substituting (\ref{eq:nutKO}) into (\ref{eq:Kout})
or (\ref{eq:Omegaout}), one obtains a nonlinear (second-order) ordinary differential equation for $\omega$, which
is deferred as we here focus on the description of the MVP.

\subsection{Modification taking into account the effects of turbulent transport}

Now, we introduce the wake modification. The essence is to enhance the turbulent transport effect, hence to decrease the eddy viscosity function towards the wake flow region that would lead to a increment of MVP. Since the original $k-\omega$ model is good at small $y$, and $\nu_T$ is generally an increasing function of $y$,
a schema to enhance the effect of turbulent transport at large $y$ is to introduce a nonlinear dependence of
$\sigma^\ast$ and $\sigma$ on $\nu_T$ such that when $\nu_T$ becomes large (close to the centerline), a new term representing
the effect of turbulent transport becomes more significant over the original setting. Specifically, we assume
\begin{equation}\label{eq:SEDsigma}
\sigma^\ast_{SED}=\sigma^\ast [1+(\gamma \nu_T)^2];\quad\quad
\sigma_{SED}=\sigma [1+(\gamma \nu_T)^2].
\end{equation}
Under this assumption, the outer similarity equations (\ref{eq:Kout}) and (\ref{eq:Omegaout}) for pipe read
\begin{eqnarray}\label{eq:SEDKout}
  {{{r^2}} \mathord{\left/
 {\vphantom {{{r^2}} {\nu _{_T}^{}}}} \right.
 \kern-\nulldelimiterspace} {\nu _{_T}^{}}} - {\beta ^*}\nu _{_T}^{}{\omega ^2} + \frac{1}{r} \frac{d}{{dr}}\left[ r{\sigma ^*}(1+\gamma^2\nu^2_T){\nu _{_T}^{}\frac{{d\left( {\nu _{_T}^{}\omega } \right)}}{{dr}}} \right] = 0;\\
\label{eq:SEDOmegaout}
  \alpha {{{r^2}} \mathord{\left/
 {\vphantom {{{r^2}} {\nu _{_T}^{2}}}} \right.
 \kern-\nulldelimiterspace} {\nu _{_T}^{2}}} - \beta {\omega ^2} + \frac{1}{r} \frac{d}{{dr}}\left[ r\sigma (1+\gamma^2\nu^2_T)
 {\nu _{_T}^{}\frac{{d\omega }}{{dr}}} \right] = 0.
\end{eqnarray}
Note that we have eliminated {the cross-diffusion term} in the $\omega$ equation (\ref{eq:Omegaout}) (i.e. $\sigma_d= 0$) as
it is unphysical here. Moreover, the rationality of (\ref{eq:SEDsigma}) can be further explained as follows. When $\gamma = 0$,
above two equations go back to the original $k-\omega$ equation;
when $\gamma > 0$, the nonlinear term is very
small in the overlap region where ${\nu _T} \approx \kappa y \to 0$, but becomes order one if $\gamma \nu_0$ is of order one
for increasing $y$ (or decreasing $r$) close to the centerline ($r \to 0$) with $\nu _{_T}^{} \to {\nu_0}$. This yields an
estimate $\gamma\sim 1/\nu_0$ (a specific value of $\gamma$ will be given below).
Note that the quadratic term ${{\gamma^2 {\nu^2 _T}}}$ in $\sigma_{SED}$ and $\sigma^\ast_{SED}$ maintains the outer similarity
of (\ref{eq:SEDKout}) and (\ref{eq:SEDOmegaout}) as the original $k-\omega$ equation does. Note also that an earlier attempt to use
a linear correction term ($\gamma {\nu _T}$) in (\ref{eq:SEDsigma}) yield a qualitatively similar result, but quantitatively
not as good as the quadratic term. {At this point, we can not show that the modification schema is unique, but the current choice in
(\ref{eq:SEDsigma}) seems to be the simplest option to give rise to satisfactory outcome. }

Substituting (\ref{eq:SEDsigma}) into (\ref{eq:Kpluspipe}) and (\ref{eq:omegapluspipe}), we obtain  a set of modified $k-\omega$
equations, which include
\begin{eqnarray}\label{eq:SEDKplus}
\quad {S^ + }{W^ + } - {\beta ^ * }{k^ + }{\omega ^ + } +\frac{1}{r}\frac{d}{{d{y^ + }}}\left[
{r\left( 1 + {\sigma_{SED} ^*}\nu_T^ +\right) \frac{{d{k^ + }}}{{d{y^ + }}}} \right] = 0\\
\alpha  \frac{{{\omega^ + }}}{{{ k^ + }}} {S^ + }{W^ + }- \beta {\omega ^ {+2} } +\frac{1}{r}
\frac{d}{{d{y^ + }}} \left[ r(1 + {\sigma_{SED} }\nu_T^ +) \frac{{d{\omega^ + }}}{{d{y^ + }}} \right] = 0\label{eq:SEDomegaplus}
\end{eqnarray}
where
$\sigma^\ast_{SED}/\sigma^\ast=\sigma_{SED}/\sigma=1+(\gamma \nu_T^+/Re_\tau)^2$.
The three equations (e.g. (\ref{eq:eddyv}), (\ref{eq:SEDKplus}) and (\ref{eq:SEDomegaplus})) form a new closed system
for the prediction of MVP in a turbulent pipe flow, on which the comparisons reported in the next subsections are based.

\begin{figure}
\subfigure[]{\includegraphics[width = 10 cm]{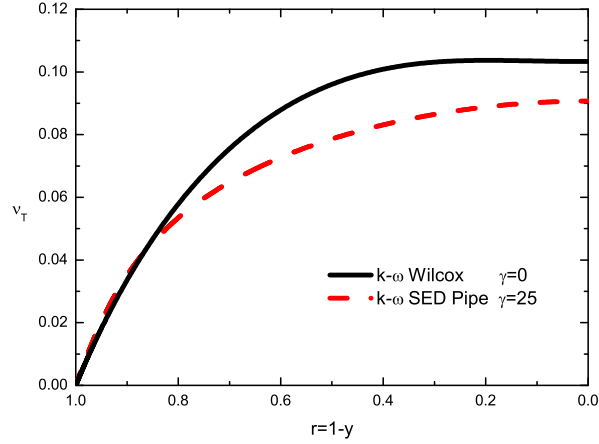}}
\subfigure[]{\includegraphics[width = 10 cm]{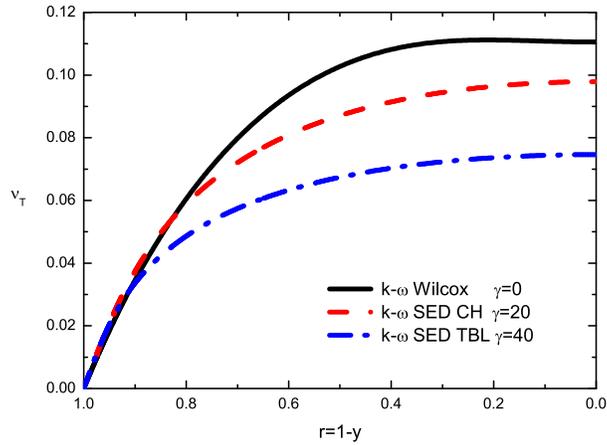}}
\caption{{Validation of eddy viscosity. (a) Pipe. Solid line: original $k-\omega$ model with $\gamma=0$; dashed line: modified model with $\gamma=25$. (b) Channel and TBL. Solid line: original $k-\omega$ model for channel; dashed line: modified model with $\gamma=20$ for channel; dash dotted line: modified model with $\gamma=40$ for TBL.}}
\label{fig:SEDnut}
\end{figure}

It is important to check if the model (\ref{eq:SEDsigma}) with a quadratic nonlinearity decreases the eddy viscosity towards centerline. The results are shown in figure \ref{fig:SEDnut}. Compared to figure \ref{fig:nut}, the model (\ref{eq:SEDsigma}) yields a smaller $\nu_0\approx 0.091$. This decrease
yields a larger mean shear near the centerline, hence a bigger increment and a stronger wake correction
away from the logarithmic profile in the mean velocity, which is required to
rectify the deficiency in the original $k-\omega$ model. Note that $\gamma$ is the only parameter to be
determined and is found to be {25} for pipe flows, independent of $Re$, which is close to $1/\nu_0$ (noting that $\gamma \nu_0\approx2.3$). Since $\gamma$ determines the amount of velocity increment beyond the log-profile, thus it is
similar to the Coles wake parameter \citep{Coles1956}. Our analysis here relates $\gamma$ to $\nu_0$, which we believe is one of the important constants in fully developed pipe turbulence. Exact relation
between $\gamma$ to $\nu_0$ needs further investigation with more general consideration of the nonlinear
eddy viscosity correction term.

\subsection{Validation of the MVP prediction by the modified $k-\omega$ equation}

\begin{figure}
\subfigure[]{\includegraphics[width = 10 cm]{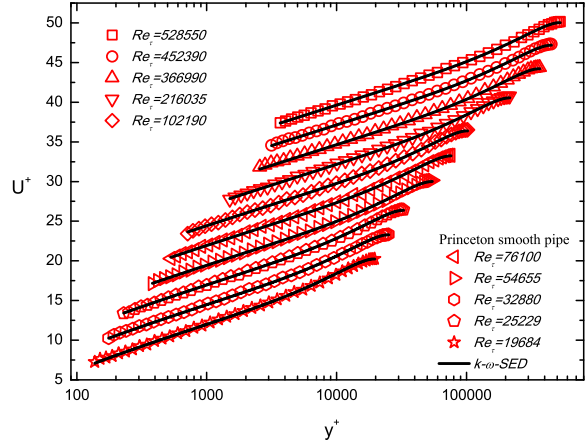}}
\subfigure[]{\includegraphics[width = 10 cm]{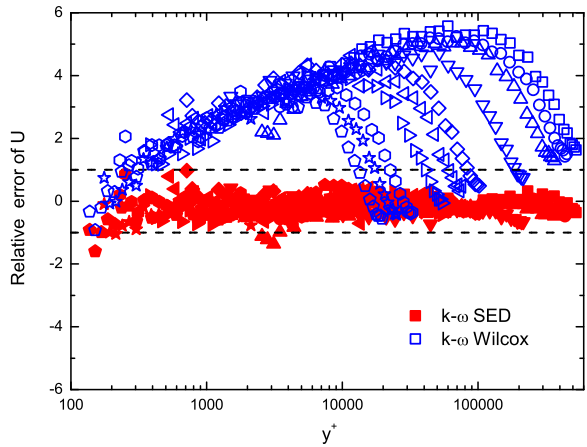}}
\caption{(a) Predictions of the modified $k - \omega $ model (lines) compared with Princeton pipe data by Zagarola \& Smits \citep{zagarola1998} for 10 MVPs. Profiles are staggered vertically for better display. (b) The relative errors, $\left( {{U^{EXP}}/{U^{Model}} - 1} \right) \times 100\% $ of the modified $k - \omega $ model (with $\kappa = 0.45$) (solid symbols), are bounded within 1\%. Also included is the Wilcox $k - \omega $ model ($\kappa = 0.40$, open symbols), showing 6\% relative error at the largest \emph{Re}.}
\label{fig:MVPe1}
\end{figure}

The modified $k-\omega $ model with $\kappa=0.45$ (i.e. $\alpha_\infty = 0.57$ and $\sigma=0.488$) and
$\gamma = 25$ (the only extra parameter) yields a significant improvement of the
prediction of the MVP for turbulent pipe. Figure \ref{fig:MVPe1}a shows
the comparison of the numerical solutions of MVPs of the modified $k-\omega $ model equations with
empirical data by Zagarola \& Smits \citep{zagarola1998} at ten different \emph{Re}'s; they are all in excellent agreement.
The relative errors, displayed in figure \ref{fig:MVPe1}b, are
uniformly bounded within 1\% - significantly smaller than that of
the original $k-\omega $ model, which shows a trend of
increase with increasing \emph{Re}, and goes up to 6\%.

\begin{figure}
\subfigure[]{\includegraphics[width = 10 cm]{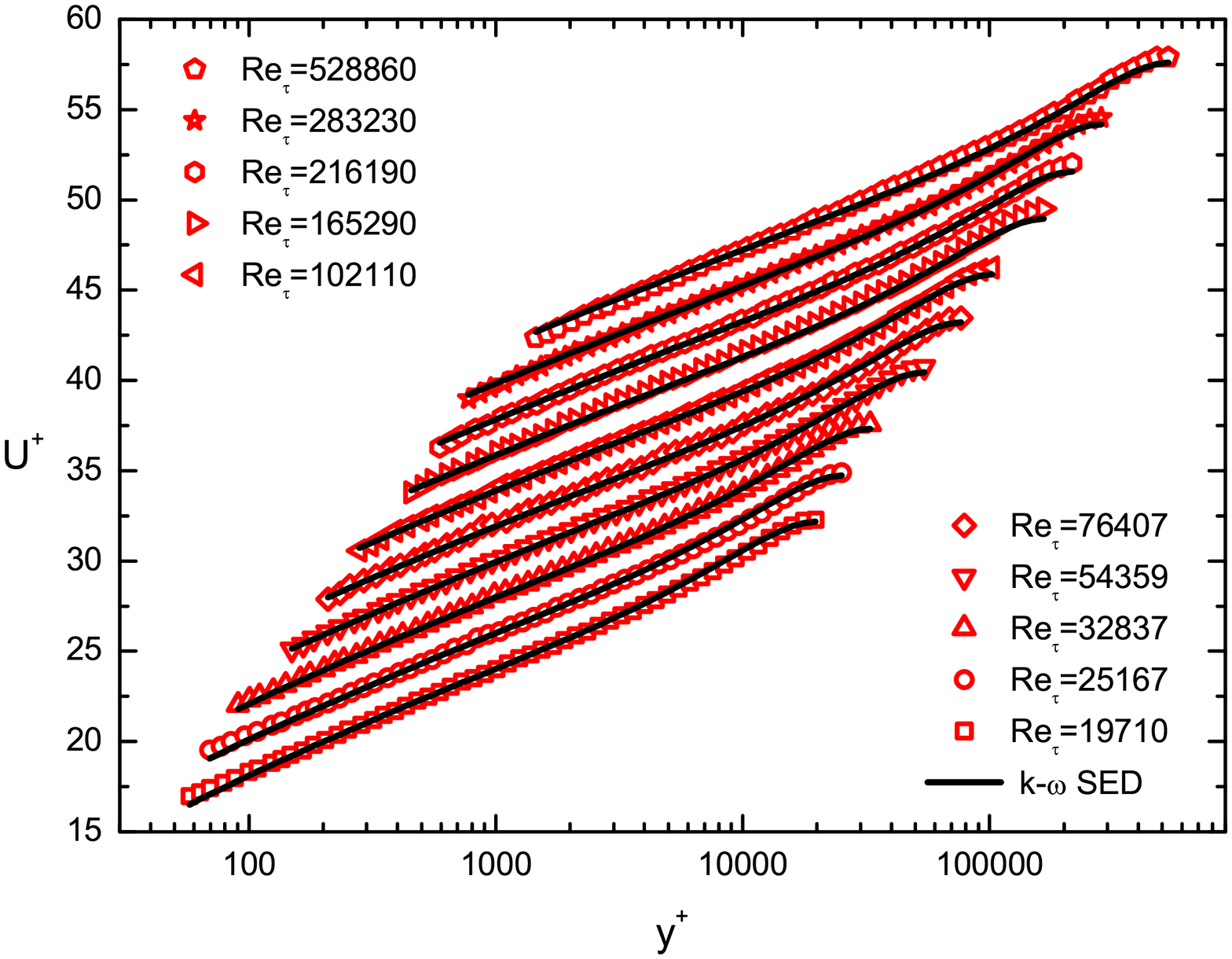}}
\subfigure[]{\includegraphics[width = 10 cm]{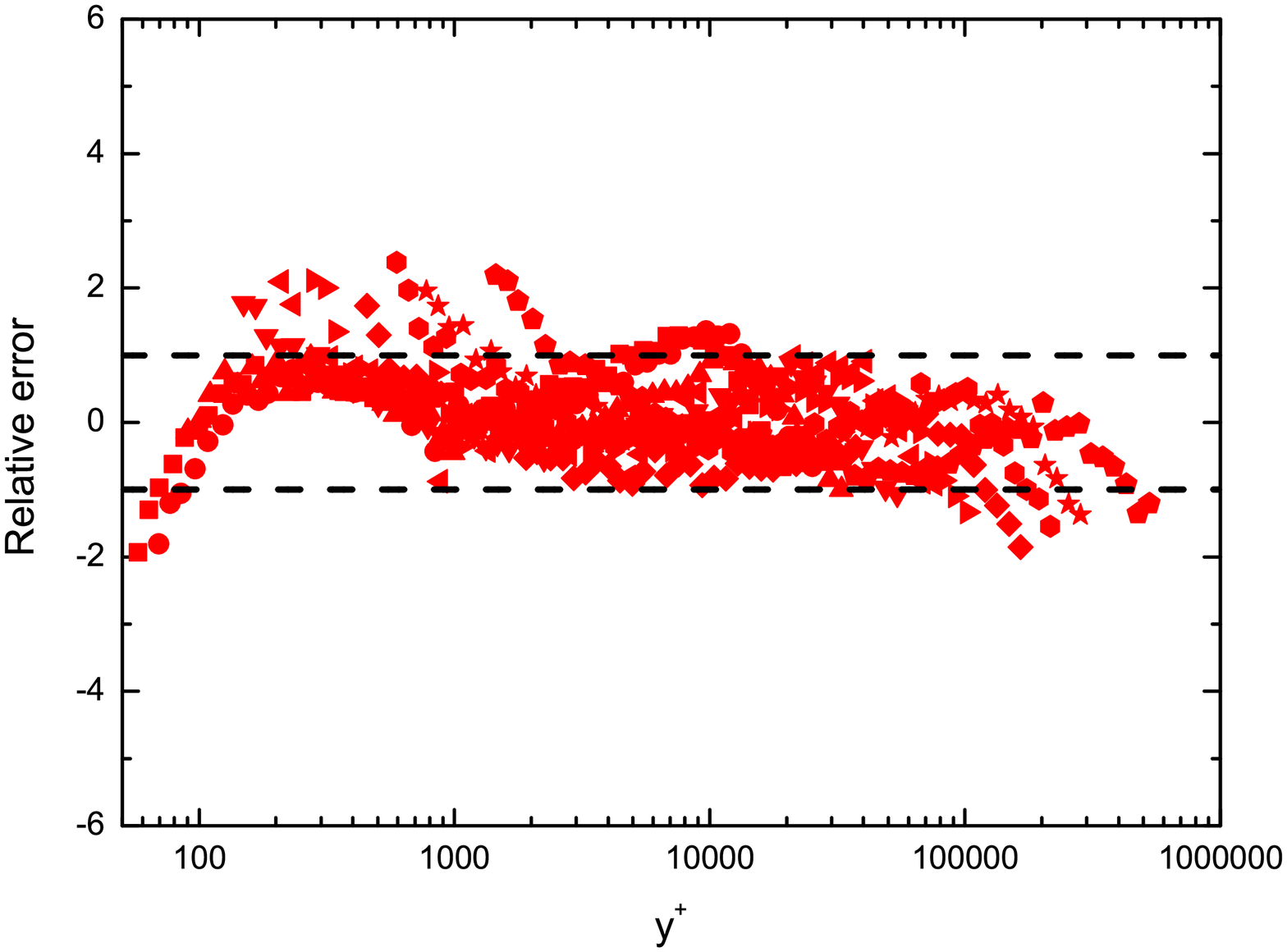}}        
  \caption{(a) Comparison of the predictions of the modified
  $k - \omega $ model (lines) with Princeton pipe data by McKeon \emph{et al} \cite{Mckeon2004} for ten MVPs.
  Each profile is staggered vertically for better display.
  (b) The relative errors of the modified $k - \omega $ model (solid symbols) are mostly bounded
  within 1\%.}
\label{fig:MSMVP}
\end{figure}

We have also compared the predictions with data by McKeon \emph{et al} \cite{Mckeon2004},
keeping the same $\kappa=0.45$ and $\gamma=25$.
The results (figure \ref{fig:MSMVP}a) are also in good agreement, with relative errors bounded within 1\% (figure \ref{fig:MSMVP}b),
while the original $k - \omega $ model over-predicts by 5\% (not shown). Note that the largest deviations are about
2\% for the first few measurement points at high \emph{Re}'s, close to experimental uncertainty. Incidentally, more recent MVP data by Hultmark \emph{et al} \citep{Hultmark2012} show noticeable deviations from earlier MVP data in
\citep{zagarola1998} and \cite{Mckeon2004}; and our study shows that
$\kappa\approx0.42$ gives a satisfactory description of the MVP data in \citep{Hultmark2012}. Particular explanation for this deviation (due to different data uncertainty) is not apparent and not discussed here.

\begin{figure}
\subfigure[]{\includegraphics[width = 10 cm]{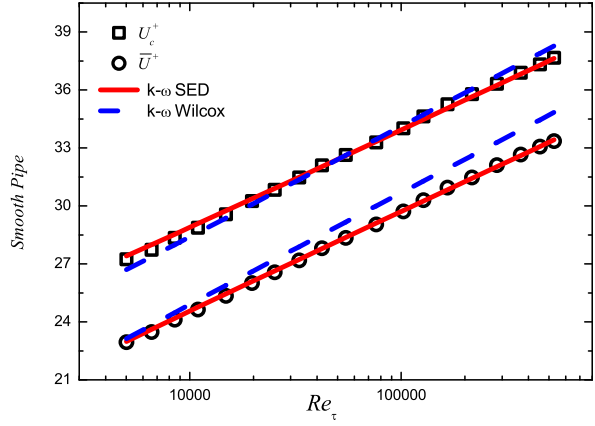}}
\subfigure[]{\includegraphics[width = 10 cm]{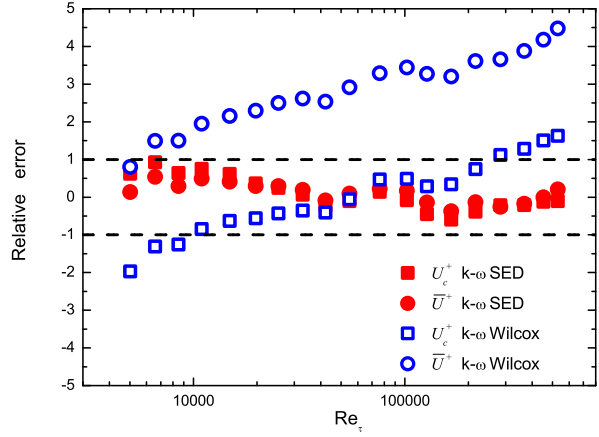}}
  \caption{(a) Predictions of the centerline velocity (square) and the volume averaged mean flux (circle)
  of the Wilcox (dashed lines) and modified (solid lines) $k - \omega $ models compared with Princeton pipe data \citep{zagarola1998}.
  (b) The relative errors (times 100) of the modified $k - \omega $ model (with  $\kappa = 0.45$ ) (solid symbols), bounded within
  1\%; Wilcox $k - \omega $ model ($\kappa = 0.40$, open symbols) shows 5\% relative error at the largest \emph{Re}.}
\label{fig:UcUb}
\end{figure}

Figures \ref{fig:UcUb}a shows the comparison of two integral quantities of engineering interest,
the centerline velocity $U^+_c$ and the volume averaged mean flux $\overline{U}^+=\int^1_0{2rU^+dr}$,
for $Re_\tau$ from $5000$ to $528550$. The modified $k-\omega$ model (solid lines) improves
significantly the original $k-\omega$ model (dashed lines), clearly demonstrated by the plot of the relative errors
in figures \ref{fig:UcUb}b. Note that the systematic bias of the original $k-\omega$ model reflects a
too small $\kappa$ used in the original $k-\omega$ model.

\begin{figure}
\subfigure[]{\includegraphics[width = 10 cm]{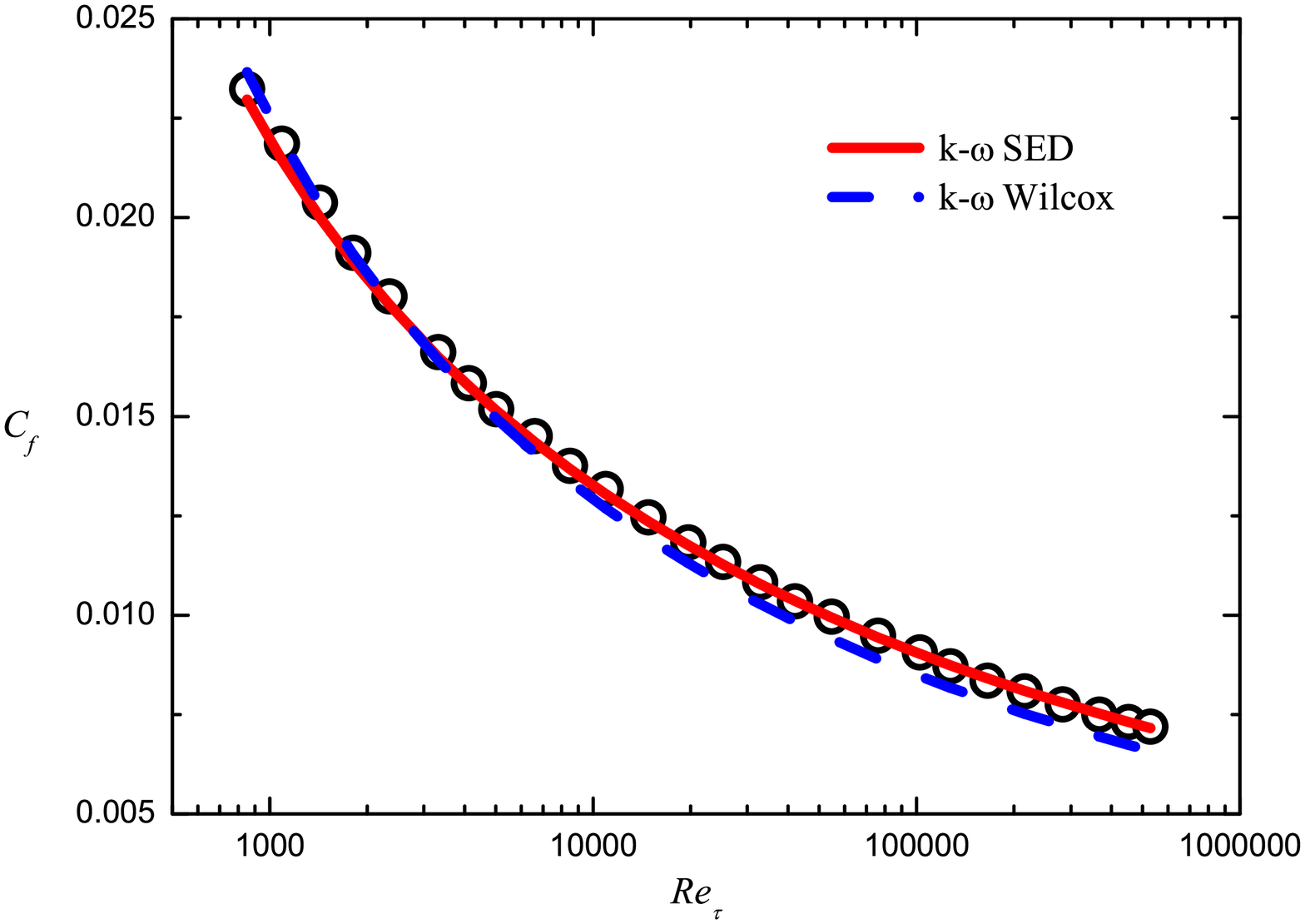}}
\subfigure[]{\includegraphics[width = 10 cm]{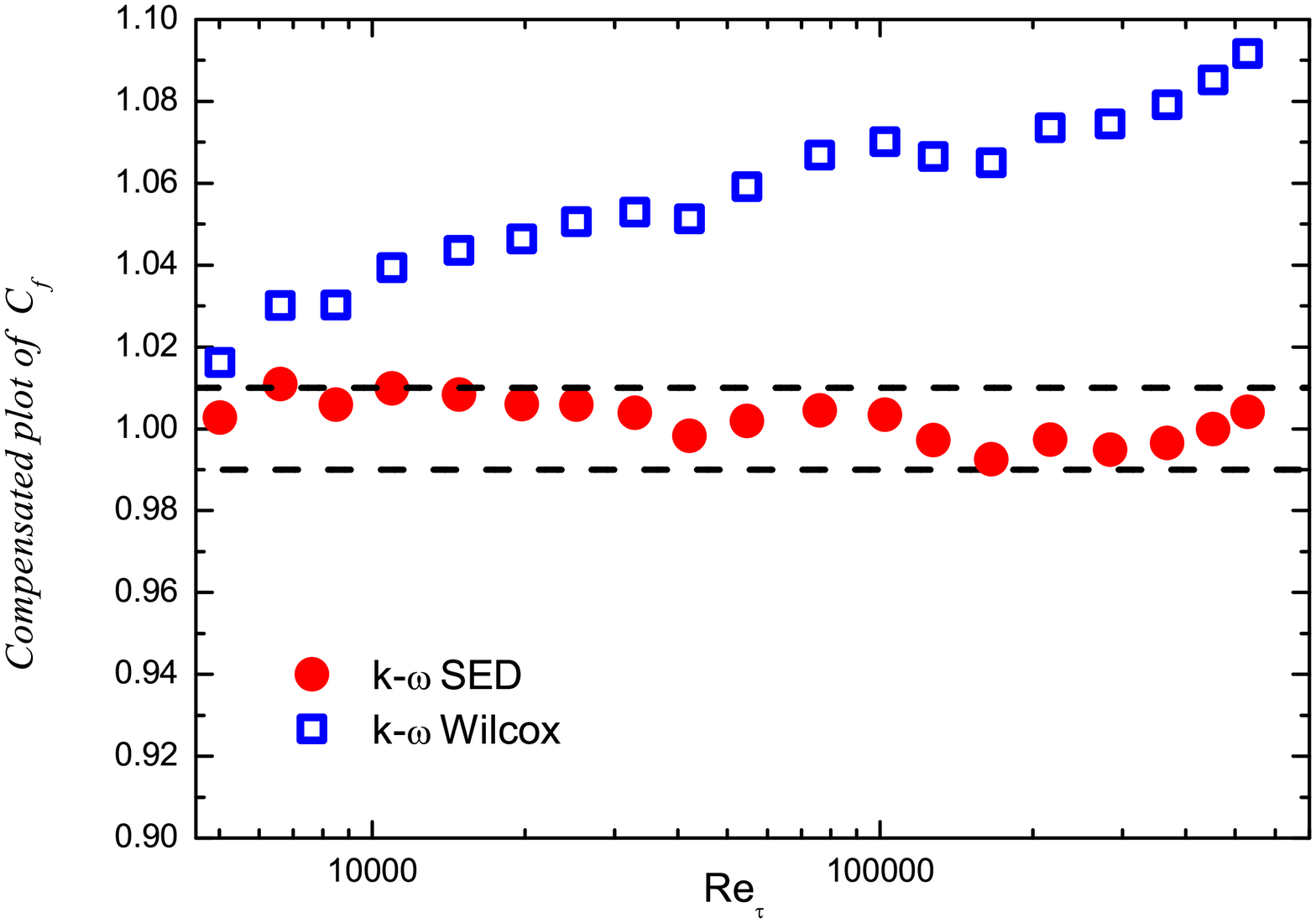}}
  \caption{(a) Predictions of the friction coefficient of the Wilcox (dashed lines) and modified (solid lines)
  $k - \omega $ models compared with Princeton pipe data \citep{zagarola1998}.
  (b) The compensated plot against the predictions of the friction coefficient of the Wilcox ($\kappa = 0.40$, open squares) and
  modified ($\kappa = 0.45$, filled circles) $k - \omega $ model. The improvement is up to 10\% at the largest $Re$.}
\label{fig:Cf}
\end{figure}

The friction coefficient is the most important quantity to predict. It is defined as $C_f=8/\overline{U}^{+2}$, and the
results are shown in
figure \ref{fig:Cf}. Note that the original $k-\omega$ model displays a trend of overestimating $C_f$ as $Re$ increases,
which reaches up to 10\% at the highest $Re$; the new model
bounds the relative errors within 1\%. This significant improvement is primarily due to the use of
the new Karman constant ($\kappa=0.45$), which affects the global trend of variation.

\subsection{{Predictions} for channel and TBL flows}

\begin{figure}
\subfigure[]{\includegraphics[width = 10 cm]{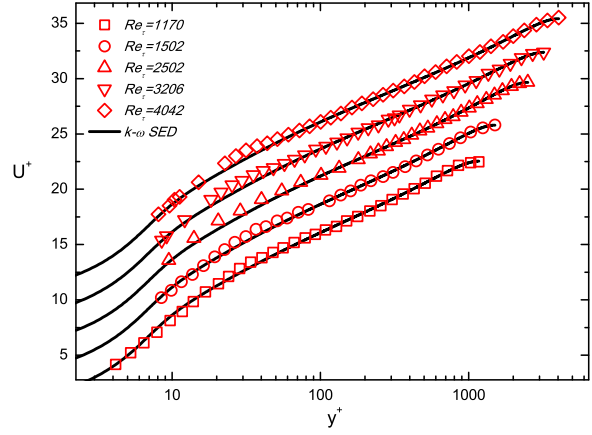}}
\subfigure[]{\includegraphics[width = 10 cm]{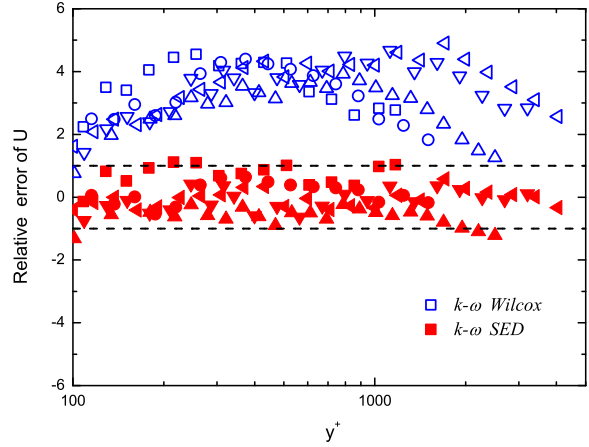}}
  \caption{(a) Predictions of five MVPs of the modified $k - \omega $ model
  (lines) compared with experimental channel flow data (symbols) by Monty \emph{et al} \cite{Monty2009}.
  Profiles staggered vertically for better display.
  (b) The relative errors, $\left( {{U^{EXP}}/{U^{Model}} - 1} \right) \times 100\% $
  of the modified $k-\omega $ model (with  $\kappa = 0.45$ ) (solid symbols), are bounded within
  1\%; Wilcox $k - \omega $ model ($\kappa = 0.40$, open symbols) shows up to 5\%
  relative error at the largest \emph{Re}.}
\label{fig:MVPCH}
\end{figure}

\begin{figure}
\subfigure[]{\includegraphics[width = 10 cm]{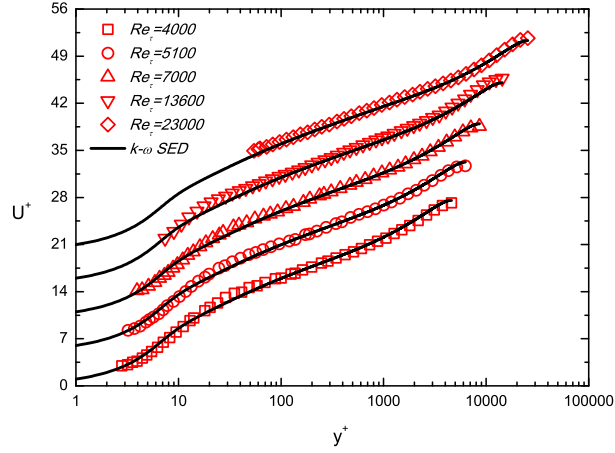}}
\subfigure[]{\includegraphics[width = 10 cm]{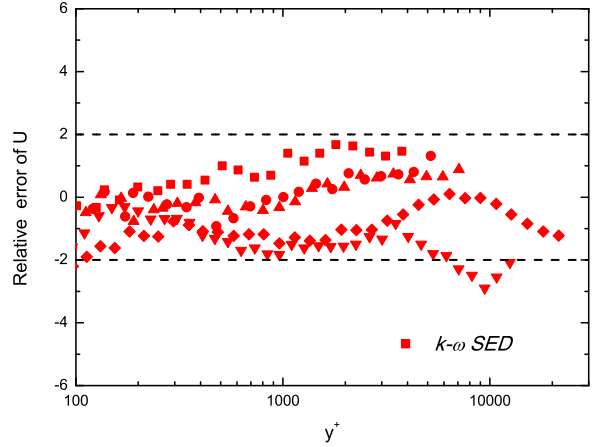}}
  \caption{(a) Predictions of five MVPs of the modified $k - \omega $ model
  (lines) compared with experimental TBL data. $Re_\tau=4000, 5100, 7000$ are from Carlier \& Stanislas \cite{Carlier2005};
  13600 from Hutchins \emph{et al} \cite{Hutchins2009}, 23000 from Nickels \emph{et al} \cite{Nickels2007}).
  Profiles are staggered vertically for better display.
  (b) The relative errors, $\left( {{U^{EXP}}/{U^{Model}} - 1} \right) \times 100\% $
  of the modified $k-\omega $ model (with  $\kappa = 0.45$ ) (solid symbols), are bounded within
  2\%.}
\label{fig:MVPTBL}
\end{figure}

Now, we discuss the extension of the model for the prediction of MVP in channel and TBL flows.
For channel flow in Cartesian coordinates, (\ref{eq:SEDKplus}) and (\ref{eq:SEDomegaplus}) are simplified to:
\begin{eqnarray}\label{eq:SEDKplusTBL}
\quad {S^ + }{W^ + } - {\beta ^ * }{k^ + }{\omega ^ + } +\frac{d}{{d{y^ + }}}\left[ {\left( 1 + {\sigma_{SED} ^*}v_T^ +\right) \frac{{d{k^ + }}}{{d{y^ + }}}} \right] = 0;\\
\alpha  \frac{{{\omega^ + }}}{{{ k^ + }}} {S^ + }{W^ + }- \beta {\omega ^ {+2} } + \frac{d}{{d{y^ + }}} \left[ (1 + {\sigma_{SED} }v_T^ +) \frac{{d{\omega^ + }}}{{d{y^ + }}} \right] = 0,\label{eq:SEDomegaplusTBL}
\end{eqnarray}
where $\sigma^\ast_{SED}/\sigma^\ast=\sigma_{SED}/\sigma=1+(\gamma \nu_T^+/Re_\tau)^2$ are the same. It is clear that the outer wake flow of channel is different from that of pipe, due to
the variation of circular to flat plate geometry. We thus introduce a different $\gamma=20$ for capturing this difference,
which corresponds to a somewhat less pronounced wake (the resulted eddy viscosity is shown in figure \ref{fig:SEDnut}b). On the other hand, we keep
the Karman constant to be the same, and $\alpha_\infty=0.52$ (hence $\sigma=0.395$) the same as the original $k-\omega$ model.
Such a different $\alpha_\infty$ between channel and pipe seems to reflect a geometry change. This schema is successfully tested against experimental channel flow data by Monty \emph{et al} \cite{Monty2009} with ${Re}_\tau$ varying from 1000 to 4000.
The predictions are shown in figure \ref{fig:MVPCH}, with again very good agreement with experiments
(i.e. relative errors uniformly bounded within 1\% for $y^+$ above 100). Note that
the original $k-\omega $ model shows an error up to 5\%.

{Here, MVP for TBLs is also predicted using the same equations as for channels.
The only change is by setting {$\gamma = 40$}}
(corresponding to a much stronger wake structure with a smaller eddy viscosity, see figure \ref{fig:SEDnut}b), while keeping $\kappa=0.45$, $\alpha_\infty=0.52$ (and $\sigma=0.395$) the same as in channels. As shown in figure \ref{fig:MVPTBL}, the relative errors are also uniformly bounded within 2\%
for $y^+$ above 100 - the same level of data uncertainty. Note that the current $k-\omega$ model does not
involve a streamwise development, so should only be considered as a mere test of how the MVPs across three
canonical wall-bounded flows (with different wakes) may be universally described with a single $\kappa=0.45$
and one varying parameter $\gamma$, up to the
data uncertainty. {Also note that the cross-diffusion term is set to zero in our model (i.e. $\sigma_d=0$).
A non-zero $\sigma_d$ may modify the prediction of MVP;
however, our study shows that complicated and unnecessary adjustments of several other parameters would be requried and hence not considered. }

\section{Modification of the $k - \omega $ equation for SMKP}\label{appA}

In Section II, we show that the original $k-\omega$ equation assumes a constant kinetic energy in the overlap
region, {called as the Bradshaw-like constant here}, which is however against recent measurements
showing an outer peak in the SMKP. Clearly, more physical consideration is needed. In our recent work \citep{Chen2015PRE},
a meso-layer is assumed to possess an anomalous dissipation triggered by an interaction between wall-attached eddies
and isotropic turbulence, which yields an anomalous scaling in the ratio of kinetic energy and
Reynolds stress (i.e. $\theta^2=K^+/W^+$). In the overlap region where $W^+\approx 1$, it implies an
anomalous scaling in the kinetic energy, and thus a modification of the {Bradshaw-like} constant.
This understanding brings in a modification on the dissipation terms in the $k-\omega$ equation, resulting
in significant improvement on the SMKP predictions. Below, we describe in detail this modification of the $k-\omega$ model
and discuss why the modification works.

\subsection{Anomalous dissipation in the meso-layer}

In our previous work \citep{she2009, Chen2015PRE}, an integrated theory of mean and fluctuation fields is proposed by the consideration of the dilation symmetry of
length functions, which play a similar role as order parameters in
Landau's mean field theory \citep{kadanoff2009more}. Specifically, two lengths are defined from a dimensional analysis
involving three physically relevant quantities in wall-shearing turbulence, i.e. $S^+$, $W^+$ and $K^+$, as:
\begin{eqnarray}\label{eq:length}
\ell^+_{12}=\sqrt{W^+}/S^+,\quad\quad \ell^+_{11}=\sqrt{K^+}/{S^+}
\end{eqnarray}
which are called {stress and energy} length functions, respectively; $1,2$ denote streamwise and normal direction, respectively.
Then, the most important energy processes, namely turbulent production $P$ and dissipation $\epsilon$, are
\begin{eqnarray}\label{eq:dissipation}
P^+=S^+W^+=\ell^{+2}_{12} S^{+3},\quad\quad \varepsilon^+=e_0\ell^{+2}_{11} S^{+3},
\end{eqnarray}
following the definition of $\ell^+_{12}$, with an additional assumption that $\ell^+_{11}\approx e_0\ell^+_{T}$
where $\ell^+_{T}$ is the usual Taylor length: $\varepsilon^+=K^{+3/2}/\ell^+_T$
(with a dimensionless coefficient $e_0$ of order one). This description yields a specific formulation of Townsend's
hypothesis \citep{Townsend1976} that momentum and kinetic energy are transferred in an analogous way
by wall attached-eddies of different length scales (see also \cite{Perry1986}).

It is easy to show that the ratio function $\theta$
exactly corresponds to the ratio of the two length functions:
\begin{equation}\label{eq:theta}
\theta\equiv\sqrt{K^+/W^+}=\ell^+_{11}/\ell^+_{12}=e_0\varepsilon^+/P^+.
\end{equation}
This relation indicates that the degree of eddy elongation (in the streamwise and normal
directions) specified by $\theta$ determines the ratio of production and dissipation. Note that if $\ell^+_{11}\sim \ell^+_{12}\approx \kappa y^+$
(in the overlap region, corresponding to the celebrated log-law: $S^+\approx 1/\ell^+_{12}\approx 1/(\kappa y^+)$),
then $\theta$ is constant. However, as mentioned in Sect. II, empirical data indicate that $\theta$ has a notable
departure from constancy in the bulk region for large (but not infinite) $Re$'s, whose modification is presented as below.

Chen \emph{et al} \cite{Chen2015PRE} introduced an anomalous scaling in the energy length
$\ell^+_{11}\propto y^{+(1+\gamma')}$, while keeping the normal scaling in the stress length
$\ell^+_{12}\propto y^{+}$. This choice amounts to keep the same production while introducing an anomaly in the dissipation. Moreover, empirical data indicate that $\theta\propto y^{+\gamma_b}$
for $y^+\ll y^+_M$, and $\theta\propto y^{+\gamma_m}$ for $y^+\gg y^+_M$, with $\gamma_b\approx 0.05$ and $\gamma_m\approx -0.09$ for large $Re$'s, where
$y^+_M=\sqrt{Re_\tau/\kappa}$ is the meso-layer thickness defined by the peak of Reynolds shear stress. The two exponents
are $Re$-independent empirical parameters, selected to fit Princeton pipe flow data \citep{Hultmark2012}, whose universality will
have to be examined against more data in the future. At this stage, a comprehensive expression for
$\theta$ valid for both the meso-layer and the bulk region is proposed as:
\begin{equation}\label{theta:formula}
{\theta} = c y^{+\gamma_b} {\left[ {1 + {{\left( {{{{y^ + }}}/{{y_M^ + }}} \right)}^2}}
\right]^{\frac{{{\gamma _m-\gamma_b}}}{2}}},
\end{equation}
where $c$ is the only adjustable $Re$-dependent parameter. According to the definition (\ref{eq:theta}),
$K^+ = W^+\theta^2\approx r\theta^2$ (since in the bulk region, $W^+\approx r$); thus (\ref{theta:formula}) fully
specifies the profile of $K^+$ in the bulk flow, and the fitting constant $c$ is fully determined by
the magnitude of the outer peak. This description was shown to agree very well with data, see
\cite{Chen2015PRE}.
Below, we will show how this yields a significant modification of the $k-\omega$ model.

\subsection{Modification by the anomalous dissipation factor}

The success in describing the $K^+$ profile in the outer region with an anomalous dissipation inspires us to
modify the dissipation term in the $k-\omega$ equation.
Let us first rewrite (\ref{eq:Kpluspipe}), (\ref{eq:omegapluspipe}) and (\ref{eq:eddyv}) as:{
\begin{eqnarray}
{S^ + }{W^ + } - \varepsilon _k^ +   + \frac{1}{r^j}\frac{d}{{d{y^ + }}}\left[r^j(1+{\sigma^\ast_{SED} }\nu_T^+) \frac{{d{k^ + }}}
{{d{y^ + }}}\right] = 0\label{eq:Kplus2}\\
\alpha {S^ + }{W^ + } - \varepsilon _\omega ^ +  + \frac{1}{r^j}\frac{k^+}{\omega^+}\frac{d}{{d{y^ + }}}\left[r^j(1+{\sigma_{SED} }\nu_T^+)
\frac{{d{\omega^ + }}}{{d{y^ + }}}\right] = 0\label{eq:Omegaplus2}\\
\frac{{{W^ + }}}{{{S^ + }}} = \nu_T^ +   = \left( {\frac{{{\alpha ^*}}}{{{\beta ^*}}}}
\right)\frac{{\varepsilon _k^ + }}{{{\omega ^{ + 2}}}} = \left( {\frac{{{\alpha ^*}}}{\beta }} \right)\frac{{\varepsilon _\omega ^
+ }}{{{\omega ^{ + 2}}}}\label{eq:nutplus2}\\\label{diss}
\varepsilon _k^ + ={\beta ^ * }{k^ + }{\omega ^ + };\quad\quad \varepsilon _\omega ^ +  = \beta{k^ + }{\omega ^ + }.
\end{eqnarray}}
where $j=0$ for channels and TBLs (Cartesian coordinate) and $j=1$ for pipes (cylindrical coordinate); the wake modification $\sigma^\ast_{SED}/\sigma^\ast=\sigma_{SED}/\sigma=1+(\gamma \nu_T^+/Re_\tau)^2$
for MVP has also been included here.

Now, we introduce modified dissipation terms to incorporate an anomalous effect by changing (\ref{diss}) as
\begin{equation}\label{dissnew}
\varepsilon _k^ + /{\beta ^ * } = \varepsilon _\omega ^ + /\beta  = {k^ + }{\omega ^ + }/\eta^2.
\end{equation}
When $\eta= 1$, (\ref{dissnew}) is the same as (\ref{diss}) representing the original $k-\omega$ equation; while $\eta\ne1$, the anomalous effect is taken into consideration as below.

According to (\ref{eq:nutplus2}), $\omega^+=\sqrt{{\alpha^\ast \varepsilon _k^+}/({\beta^\ast\nu_T^+})}$.
Substituting it into (\ref{dissnew}) yields $\eta^2=k^+\sqrt{\alpha^\ast\beta^\ast /(\varepsilon _k^+ \nu_T^+)}$.
In the overlap region, $\alpha^\ast\approx1$, $\beta^\ast\approx\beta_0^\ast=0.09$, and $ \varepsilon _k^ +\approx {S^ + }{W^ + }$
with $S^ +\nu^+_T=W^+$. We thus have
\begin{equation}\label{eq:etaA}
\eta^2\approx({k^+/W^+})\sqrt{\beta^\ast_0}.
\end{equation}
Finally, substituting (\ref{eq:theta}) and
(\ref{theta:formula}) into (\ref{eq:etaA}), we obtain $\eta\approx \theta ({\beta^\ast_0}/4)^{\frac{1}{4}}$, namely
\begin{equation}\label{eq:etaB}
\eta\approx  c({\beta^\ast_0}/4)^{1/4} y^{+\gamma_b}
{\left( {1 + {{\left( {\frac{{{y^ + }}}{{y_M^ + }}} \right)}^2}} \right)^{\frac{{{\gamma_m-\gamma_b}}}{2}}}.
\end{equation}
(\ref{eq:etaB}) is not yet entirely satisfactory, because it leads to
$\eta\rightarrow 0$ as $y^+\rightarrow0$, if $\gamma_b>0$.
A further modulation on
$y^{+\gamma_b}$ can be introduced, in accordance with our multi-layer ansatz, yielding:
\begin{equation}\label{lamda3}
\eta = c'{\left( {1 + {{\left( {\frac{{{y^ + }}}{{y^+_B }}} \right)}^2}} \right)^{\frac{{{\gamma_b}}}{2}}}
{\left( {1 + {{\left( {\frac{{{y^ + }}}{{y_M^ + }}} \right)}^2}} \right)^{\frac{{{\gamma_m-\gamma_b}}}{2}}}
\end{equation}
with $c'=(y^+_B)^{\gamma_b}c (\beta_0^*/4)^{\frac{1}{4}}$ and $y^+_B=40$ indicating the buffer layer thickness.
It can be checked that (\ref{lamda3}) satisfies
the near wall condition $\eta\rightarrow c'$ (which is order 1) as $y^+\rightarrow0$ for $y^+\ll y^+_B$, as well as
the overlap region condition
(\ref{eq:etaB}) for $y^+\gg y^+_B$. This is the final form of the anomalous scaling modification of the $k-\omega$ equation.

\begin{figure}
\subfigure[]{\includegraphics[width=10.cm]{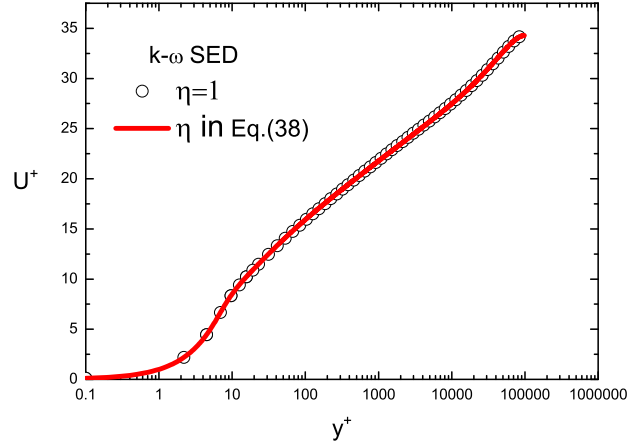}} 
\subfigure[]{\includegraphics[width=10.cm]{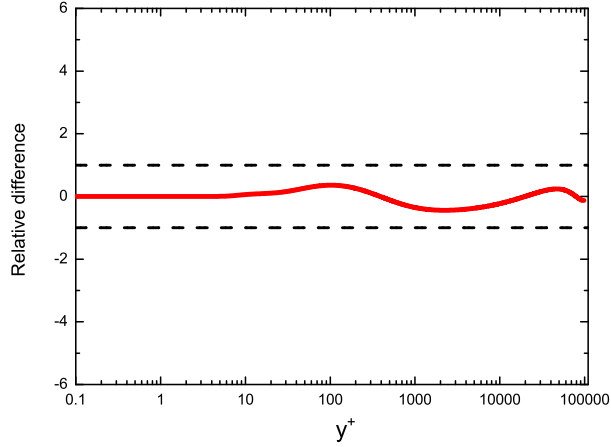}}
\caption{(a) Predicted mean velocities with and without anomalous dissipation modification in (\ref{dissnew}).
Note that $\eta=1$ (symbols) means no modification while $\eta$ in (\ref{lamda3}) (solid line) indicates the anomalous dissipation modification. (b) Relative differences (times 100) between the two mean velocity profiles in (a) are uniformly bounded within 0.4\%, indicating that the $\eta$ modification affects MVP very little.}
\label{fig:EtaMVP}
\end{figure}

Before comparison with data, it is necessary to check that the modifications (\ref{dissnew}) and (\ref{lamda3}) have negligible influence
on the prediction of MVP - otherwise we need to revise the previous modification. This is indeed verified, as shown in
figure \ref{fig:EtaMVP},
where $\eta$ in (\ref{lamda3}) changes the MVP by no more than 0.4\%. This is
tiny, compared to experimental uncertainty. Hence, (\ref{eq:Kplus2}), (\ref{eq:Omegaplus2}), (\ref{eq:nutplus2})
and (\ref{dissnew}) (with (\ref{lamda3})) are our final expressions for the modified $k-\omega$ equation, which results in accurate
predictions of both MVP and SMKP, as demonstrated below.

\subsection{Comparison of the new $k-\omega$ equation prediction with data}

\begin{figure}
\begin{center}
\subfigure[]{\includegraphics[width=10.cm]{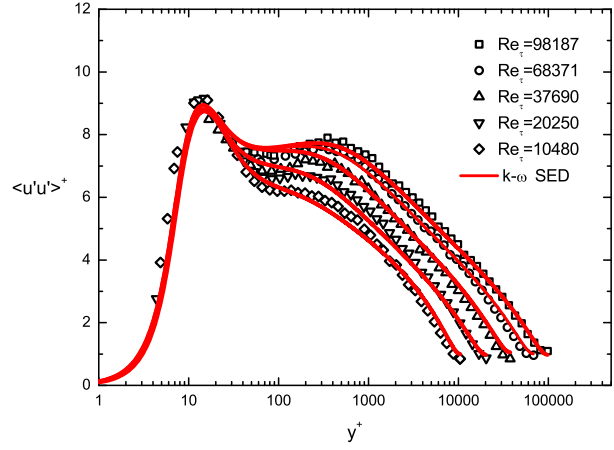}}
\subfigure[]{\includegraphics[width=10.cm]{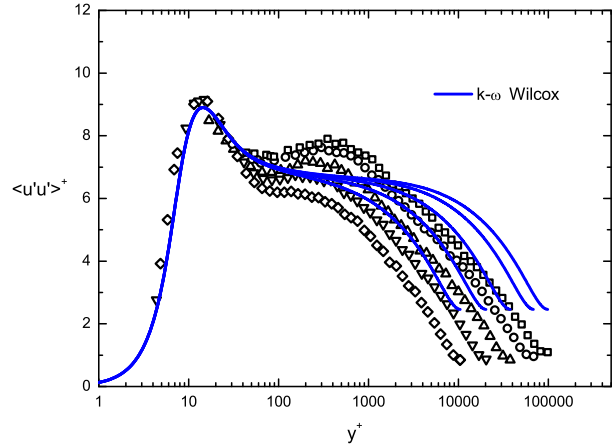}}
\caption{Comparison between {Princeton pipe data (symbols) \citep{Hultmark2012}}
and model predictions (lines). (a) Modified $k-\omega$ equation using (\ref{lamda3}); (b) Original $k-\omega$
equation with $\eta=1$.}
\label{fig:MKP}
\end{center}
\end{figure}

\begin{figure}
\begin{center}
\subfigure[]{\includegraphics[width=10.cm]{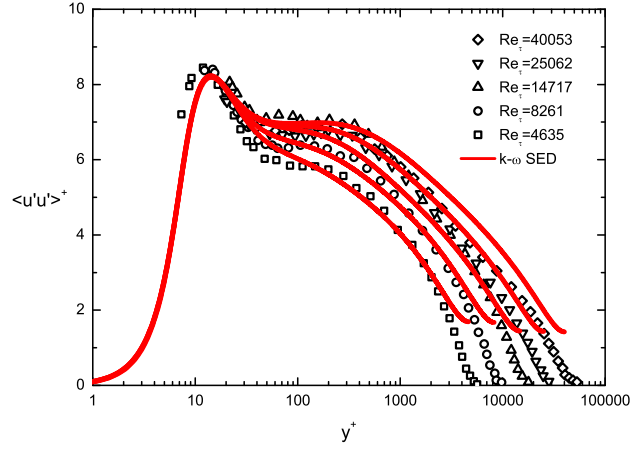}}
\subfigure[]{\includegraphics[width=10.cm]{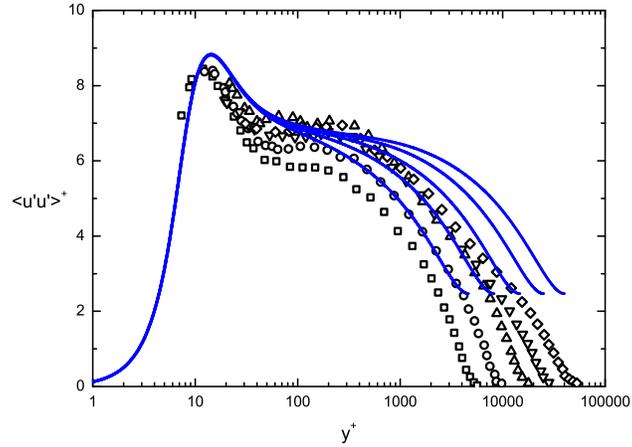}}
\caption{Comparison between Princeton TBL data (symbols) \citep{Vallikivi2015jfm}
and model predictions (lines). (a) Modified $k-\omega$ equation using (\ref{lamda3}); (b) Original $k-\omega$
equation with $\eta=1$.}
\label{fig:MKPTBL}
\end{center}
\end{figure}

Figures \ref{fig:MKP} a, b show the predictions of SMKP in pipes by the modified
and the original $k-\omega$ equations, respectively. Data are from Hultmark \emph{et al} \cite{Hultmark2012} for five different Re's. The modified model reproduces the outer peak in sharp contrast to the original model predicting a constant plateau in the overlap region.
The improvement is significant, entirely due to the introduced anomalous scaling factor (\ref{lamda3}), with
parameters given as below. The two scaling exponents are set to be constants: $\gamma_b=0.05$ and $\gamma_m=-0.09$
(except for $\gamma_m=-0.06$ for the smallest $Re_\tau=10480$), as mentioned above;
the meso-layer thickness is fully specified as $y^+_M=\sqrt{Re_\tau/\kappa}$
with $\kappa=0.45$. Finally,  $c'$ is a $Re$-dependent parameter,
being set as 0.92, 0.97, 1.0, 1.0 and 1.0, respectively, for the five Re's varying from $Re_\tau=10480$ to $Re_\tau=98187$;
thus, $c'$ increases with increasing $Re$'s and saturates to 1.
Note that as shown in \cite{Chen2015PRE}, $c'<1$ would under predict the inner peak magnitude,
because $\eta\rightarrow c'$ as $y^+\rightarrow0$. So, for the two moderate $Re$'s, to compensate for the influence
by $c'<1$ for the inner peak,
two parameters in the original $k-\omega$ model, i.e. $R_k$ and $R_\beta$ (determining the transitions of $\alpha^\ast$
and $\beta^\ast$, respectively) are calibrated empirically. The original values, $R_k=6$ and $R_\beta=8$,
are now set to be $R_k=7$ and $R_\beta=11$ for $Re_\tau=10480$; and $R_k=6$ and $R_\beta=9$ for $Re_\tau=20250$,
to take into account the moderate $Re$ effect. The results are shown in figure \ref{fig:MKP}a,
where the resulted inner peaks are invariant for all $Re$'s, in satisfactory agreement with empirical data.
In summary, with slight $Re$-dependence of $R_k$ (going from 7 to 6) and $R_\beta$ (going from 11 to 9 and then to 8),
we have obtained a consistently good description of SMKP
in turbulent pipes through the factor $\eta$ in (\ref{lamda3}) over a wide range of $Re$'s.

We now apply the above description of the SMKP for TBL
(channels are not discussed here for the lack of high \emph{Re} data), in the hope of validating
a unified $k-\omega$ model for all three canonical wall flows. Thanks to the recent measurements of SMKP by Vallikivi \emph{et al} \cite{Vallikivi2015jfm}
of high \emph{Re} TBL flows, one recognizes a similar outer peak and a local logarithmic profile as
in pipes; in contrast, the inner peak is slightly lower (by $0.5u^2_\tau$) than that in pipes.
The experimental profiles are shown in figure \ref{fig:MKPTBL} for five different $Re_\tau$'s
varying from 4635 to 40053 (over nearly one decade). We set $j=0$ in
(\ref{eq:Kplus2})-(\ref{eq:nutplus2}) for flat geometry ($\kappa$, $\alpha_\infty$, $\sigma$ and $\gamma$ are the
same as in figure \ref{fig:MVPTBL}). Comparison is made against $\eta=1$ in (\ref{dissnew}),
the original $k-\omega$ model, which, as shown in figure \ref{fig:MKPTBL}b,
yields a nearly constant plateau $\left<u'u'\right>^+\approx6.6$ in the bulk flow region, significantly away from
data. Figure \ref{fig:MKPTBL}a shows the modified $k-\omega$ model with (\ref{lamda3}). The parameters are given as below.
First, $\gamma_b=0.05$, $y^+_B=40$ and $R_k=6$ are the same for all $Re$'s. For the two smallest $Re$'s, i.e.
$Re_\tau=4635$ and $8261$, $\gamma_m=-0.03$, $-0.04$; $c'=0.92$, $0.94$; and $R_\beta=9$, $8$, respectively.
For the remaining profiles, $\gamma_m=-0.06$, $c'=0.96$ and $R_\beta=7$ are kept the same for all three larger $Re$'s.
The new predictions improve
the original ones notably in the bulk flow region as well as in the inner peak region ($y^+\approx10\sim100$),
shown in figure \ref{fig:MKPTBL}a.

{Note that the present version of the model does not apply to
two-dimensional simulation of TBL, because (\ref{eq:Kplus2})-(\ref{eq:nutplus2}) do not involve
any streawise variation along with inflow and outflow boundary conditions. The departure of predictions from data near
the freestream in figure \ref{fig:MKPTBL}a reflects this; further treatment of this issue is beyond the scope of this paper.}

\subsection{Discussion for asymptotically large $Re$'s}

An important remaining issue is the asymptotic behavior of SMKP for large $Re$'s. It is a subtle issue compared
to that of MVP which does not involve the dissipation anomaly. The question is how the anomaly varies as
$Re$ increases, since it is not certain whether current data have already reached the asymptotic similarity state
such that the parameters in (\ref{fig:MKP}) become invariant for all $Re$'s.

\begin{figure}
\subfigure[]{\includegraphics[width=10.cm]{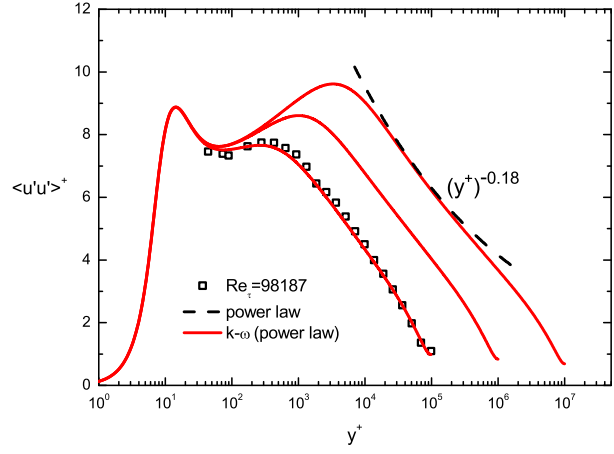}}
\subfigure[]{\includegraphics[width=10.cm]{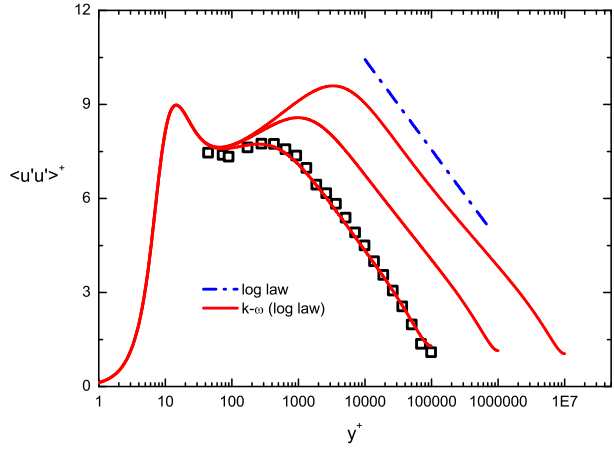}}
\centering
\caption{Comparison between data (symbols) and the predictions of the modified $k-\omega$ equation (solid lines).
(a) The three parameters are invariant as $Re$, i.e. $y^+_B=40, \gamma_b=0.05, \gamma_m=-0.09$. Dashed line show the local power law $(y^+)^{-0.18}$. (b) While $y^+_B=40$ and $\gamma_b=0.05$ are fixed, $\gamma_m$ varies
from -0.09 to -0.088, and to -0.085 for $Re_\tau=10^5, 10^6$ and $10^7$, respectively. Dashed line shows the
log distribution with the slope $-1.25$.}
\label{fig:MKPReInF}
\end{figure}

Let us first explain the asymptotic behavior of SMKP when we keep $\gamma_b$, $\gamma_m$ fixed
(the meso layer thickness $y^+_M$ is fixed to be $\sqrt{Re_\tau/\kappa}$).
In this case, one obtains a power law scaling $\theta\propto (y^+)^{\gamma_m}=(y^+)^{-0.09}$ and
hence $K^+\propto (y^+)^{2\gamma_m} \propto (y^+)^{-0.18}$ for $y^+\gg y^+_M\propto\sqrt{Re_\tau}$.
The resulting SKMP are shown in figure \ref{fig:MKPReInF}a, where the solid lines are predictions
for $Re_\tau=10^5$, $10^6$ and $10^7$. While the inner peak keeps invariant, the outer peak exceeds the inner
one at $Re_\tau=10^7$. Moreover, the dashed line indicates the local power law scaling, agreeing well with the
numerical result of the modified $k-\omega$ equation. The result indicates a possible power law scaling of SMKP
for asymptotically large $Re$'s, which is discernible from the logarithmic profile for $Re_\tau$ over $10^7$.

On the other hand, it is interesting to explore the possibility with the logarithmic scaling for SMKP in
(\ref{eq:Kplus}). In fact, (\ref{lamda3}) also can result in a satisfactory approximation
of the log profile through a Re-dependent scaling exponent, in analogy to the power law by Barenblatt \cite{Barenblatt}. Note that from (\ref{lamda3}), $\eta\propto y^{+\gamma_m}{Re_\tau}^{\frac{\gamma_b-\gamma_m}{2}}$ for $y^+\gg \sqrt{Re_\tau}$, and $K^+\propto\eta^2\propto y^{+2\gamma_m}Re_\tau^{\gamma_b-\gamma_m}$. Thus the logarithmic diagnostic function $\Gamma=y^+dK^+/dy^+\propto \gamma_m Re_\tau^{\gamma_b-\gamma_m}y^{+2\gamma_m}$. Since $\gamma_b\approx0.05$ is believed to be invariant, if $\gamma_m\propto Re_\tau^{-\gamma_b}$ so that $\gamma_m Re_\tau^{\gamma_b-\gamma_m}\rightarrow const.$ and $y^{+2\gamma_m}=\exp(2\gamma_m\ln {y^+})\rightarrow 1$ as $Re_\tau\rightarrow\infty$, then a constant $\Gamma\rightarrow const.$ is expected indicating the logarithmic scaling at asymptotically large $Re$'s. In other words, to meet the log scaling as $Re_\tau\rightarrow\infty$, we have $\gamma_m\propto Re_\tau^{-0.05}$. It turns out that the convergence to such an asymptotic scaling is very slow, and here we verify that the magnitude of $\gamma_m$ indeed decreases with increasing $Re$. In figure \ref{fig:MKPReInF}b, we keep $\gamma_b=0.05$ invariant and let $\gamma_m=-0.09, -0.088, -0.085$ for $Re_\tau=10^5, 10^6, 10^7$, respectively. The predicted SMKP are shown to exhibit more likely the log profiles marked by the dashed line, thus supporting current analysis.

Therefore, while the asymptotic scaling for SMKP is to be confirmed by more data, our modification factor (\ref{lamda3})
reveals the capability to adapt to different asymptotic states of large $Re$.

\section{Discussion and conclusion}

{We have presented a modified $k-\omega$ equation which predicts both MVP
and SMKP with better accuracy than the Wilcox $k-\omega$ model for three canonical wall-bounded turbulent flows
(channel, pipe and TBL), over a wide range of $Re'$s.} Three modifications are introduced: 1) an adjustment of the Karman constant
for better prediction of the overlap region, 2) an enhanced nonlinear
turbulent transport (with parameter $\gamma$) corresponding to the wake structure, 3) and
a dissipation factor $\eta$ exhibiting an anomalous scaling for the meso layer.
The first two modifications yield significant improvement of the MVP prediction (by nearly 10\%), agreeing well with Princeton
pipe data over a wide range of $Re$'s, and the last one reproduces the outer peak of SMKP for the first time.

Some interesting issues need further discussions here. {The most important result is that
a single $\kappa\approx 0.45$ leads to high accurate descriptions for more than 30 MVPs for all three canonical wall flows
(channel, pipe and TBL), when an appropriate wake parameter $\gamma$ is used. The original $k-\omega$ equation is shown to fail in predicting correct overlap regions and wakes,
and our modified version results in a good accuracy. Although it may not be considered as the definitive proof of the universal Karman constant since $\gamma$ is flow-dependent,
we believe that $\kappa$, being part of the universal wall function, should itself be universal.}

{Second, the $\gamma$ modification is physically driven, and it is important to
discuss its generality. The nonlinear enhancement term in the eddy viscosity $\nu_T$ is a generic feature of
the wake (increasing velocity increment above the log law), supported by the above mentioned fact that a
singly varying $\gamma$ (20, 25 and 40 for channel, pipe and TBL, respectively) produces correctly all three wake profiles.
Note that a variation of $\gamma$ from 20 to 40 increases the wake profile by $3\%$, and it may be used to predict
other more complicated wall flows with different wakes. Hence,
it will be worthwhile to test how $\gamma$ changes with pressure gradients \citep{NASA2014}.}

Third, for SMKP, it is only apparent that many parameters are involved, including
the exponents $\gamma_b$ and $\gamma_m$, and coefficient $c'$; in reality, these parameters vary little.
The coefficient $c'$ is essentially one, varying from 0.92 to 1, and
$\gamma_b\approx 0.05$ is also held fixed for both pipe and TBL. The fact $\gamma_b$ is constant is not surprising, since
it characterizes the meso layer which is part of the overlap region, and hence part of the universal wall function.
The only significant variation occurs in $\gamma_m$, -0.09 to -0.06 for pipe and TBL, respectively, with
slight $Re$-dependence; this is a typical feature of the bulk or wake property. Further study is needed to verify the
universality of these parameter values.

A conclusion drawn from the above results is that significant improvement of the accuracy of CFD turbulence models can
benefit from a careful study of the multi-layer structure of the mean flow. The present work shows that
the wake modification and the anomalous dissipation factor
for the meso-layer are good examples of physically inspired changes. Future work will investigate the connection between
the multi-layer parameters and the $k-\omega$ model coefficients (i.e. $\beta_0$, $\beta^*$, $\sigma$, etc.) for wall functions,
where the latter may eventually be replaced by the multi-layer parameters (amenable to adjustment for more complex flows).

\section*{Acknowledgement}

We thank B.B. Wei who has contributed significantly during the initial stage of the work.
This work is supported by National Nature Science Fund 11221062, 11452002 and by MOST 973 project 2009CB724100.

\appendix
{\section{Integrated Mean momentum equation for pipe flows}
For a fully-developed steady pipe flow, the dimensional mean momentum equation (in cylindrical coordinates) reads:
\begin{equation}\label{eq:A1}
-\frac{\partial P}{\partial x}+\frac{1}{\hat{r}}\frac{\partial}{\partial \hat{r}}\left[\hat{r}\left(\nu \frac{\partial U}{\partial \hat{r}}-\langle u'v'\rangle\right) \right]=0
\end{equation}where $\hat{r}=r \delta$ ($\delta$ the pipe radius). Integrating (\ref{eq:A1}) with respect to $\hat{r}$ and using the constant
pressure gradient condition (i.e. ${\partial_x P}=const.$), we obtain
\begin{equation}\label{eq:A2}
\nu \frac{\partial U} {\partial \hat{r}} - \langle u'v'\rangle = \frac{\hat{r}}{2} \frac{\partial P}{\partial x}.
\end{equation}
Defining the friction velocity as $u_\tau\equiv\sqrt{-\nu \partial_{\hat{r}} U|_{\hat{r}=\delta}} =\sqrt{-(\delta/2) \partial_x P}$,
and assigning the positive $v'$ as the wall normal velocity pointing away from the wall, (\ref{eq:A2}) is then
\begin{equation}\label{eq:A22}
\nu \frac{\partial U} {\partial y} - \langle u'v'\rangle =\hat{r} u_\tau^2= (\delta-y)u_\tau^2.
\end{equation}
Normalizing (\ref{eq:A22}) with $u_\tau$ and $\nu/u_\tau$ and defining $Re_\tau=u_\tau \delta/\nu$, we obtain the integrated mean
momentum equation (\ref{eq:MME}) for pipes - the same as for channels.}

{\section{Dimensional modified $k-\omega$ equations}
We rewrite the modified $k-\omega$ equations, i.e. equations (\ref{eq:Kplus2}) to (\ref{diss}), into dimensional formulations, i.e.
\begin{eqnarray}
-\langle u'v' \rangle \frac{\partial U}{\partial y} - {\beta ^ * }{k }{\omega } + \nonumber\\ \frac{1}{(\delta-y)^j}\frac{\partial}
{{\partial{y }}}\left[(\delta-y)^j(\nu+{\sigma^\ast_{SED} }\nu_T) \frac{{\partial{k }}}
{{\partial{y }}}\right] = 0;\label{eq:KA}\\
-\alpha \frac{\omega}{k} \langle u'v' \rangle \frac{\partial U}{\partial y} - {\beta }{}{\omega^2 } +  \nonumber\\ \frac{1}
{(\delta-y)^j}\frac{\partial}{{\partial{y }}}\left[(\delta-y)^j(\nu+{\sigma_{SED} }\nu_T)
\frac{{\partial{\omega}}}{{\partial{y }}}\right] = 0\label{eq:OA}.
\end{eqnarray}
where $j=0$ for channel and TBL and $j=1$ for pipe;  $\sigma^\ast_{SED}/\sigma^\ast=\sigma_{SED}/\sigma=1+\gamma^2 Re^{-2}_\tau
(\nu_T/\nu)^2$. Note that the eddy viscosity $\nu_T=-\langle u'v' \rangle/{\partial_y U} =  \alpha ^* k \omega $ is the same
as the original $k-\omega$ equation. Also, (\ref{eq:KA})-(\ref{eq:OA}) should be solved together with the integrated mean momentum
equation (\ref{eq:A22}).}

\bibliographystyle{ieeetr}


\end{document}